\documentclass[aps,prd,reprint,
superscriptaddress, 
nofootinbib,preprintnumbers]{revtex4-2}  
\usepackage{graphicx}  
\usepackage{dcolumn}   
\usepackage{bm}        
\usepackage{amssymb,amsfonts,amsmath,physics}   
\usepackage[dvipsnames]{xcolor}
\usepackage{slashed}
\usepackage{multirow}
\usepackage{dsfont}
\usepackage{booktabs}
\usepackage{comment}
\usepackage{inputenc}
\usepackage[bookmarks, breaklinks, colorlinks,urlcolor=blue, citecolor=red, 
linkcolor=blue]{hyperref}
\usepackage[normalem]{ulem}
\usepackage{orcidlink}
\usepackage{tikz-feynman}
\usepackage{soul}
\tikzfeynmanset{compat=1.1.0,arrow size = 1pt}
\usetikzlibrary{arrows.meta}

\tikzset{
    vertex/.style={circle, draw, minimum size=1.2em, inner sep=2pt},
    small dot/.style={circle, fill=black, minimum size=3pt, inner sep=0pt},
    edge/.style={->, >=Latex}    
}

\def\r {\rightarrow}
\def \n {\nonumber}

\def \eq {\text{eq}}
\def \vs {{\varepsilon_\Sigma}}

\newcommand{\beqa}{\begin{eqnarray}}
\newcommand{\eeqa}{\end{eqnarray}}
\def\hc {\text{H.c.}}
\newcommand{\nn}{\nonumber}

\begin{document}

\title{ 
Thermal Leptogenesis in the Babu-Nandi-Tavartkiladze Model of Neutrino Mass}

\author{P. S. Bhupal Dev $^{}{}$\orcidlink{0000-0003-4655-2866}}
\email{bdev@wustl.edu}
\affiliation{Department of Physics and McDonnell Center for the Space Sciences, Washington University, St. Louis, Missouri 63130, USA}
\affiliation{PRISMA$^{++}$ Cluster of Excellence \& Mainz Institute for Theoretical Physics, 
Johannes Gutenberg-Universit\"{a}t Mainz, 55099 Mainz, Germany}

\author{Srubabati Goswami $^{}{}$\orcidlink{0000-0002-5614-4092}}
\email{sruba@prl.res.in}
\affiliation{Theoretical Physics Division, Physical Research Laboratory, Navrangpura, Ahmedabad 380009, India}

\author{Debashis Pachhar $^{}{}$\orcidlink{0000-0001-8931-5321}}
\email{debashisp@iitk.ac.in}
\affiliation{Theoretical Physics Division, Physical Research Laboratory, Navrangpura, Ahmedabad 380009, India}
\affiliation{
Indian Institute of Technology Kanpur, Kalyanpur, Kanpur Nagar, Uttar Pradesh 208016, India}

\author{Drona Vatsyayan $^{}{}$\orcidlink{0000-0002-6868-3237}}
\email{drona@physics.carleton.ca}
\affiliation{Department of Physics, Carleton University, Ottawa, ON K1S 5B6, Canada}
\affiliation{Arthur B. McDonald Canadian Astroparticle Physics Research Institute, 64 Bader Lane, Queen's University, Kingston, ON K7L 3N6, Canada}

\begin{abstract} 
We investigate neutrino mass and thermal leptogenesis in the Babu-Nandi-Tavartkiladze (BNT) model featuring a scalar quadruplet ($\Phi$) and a pair of vector-like fermion triplets ($\Sigma$). In this framework, neutrino masses are generated via an effective dimension-7 operator $LLHH(H^{\dagger}H)/\Lambda^3$ at the tree level and via the dimension-5 operator $LLHH/\Lambda$ at the one-loop level. 
It naturally accommodates sub-eV neutrino masses even if the new physics scale $\Lambda$ is $\mathcal{O}(\rm TeV)$, thus making the model a compelling target for experimental searches. We explore the viability of thermal leptogenesis in this model, which is distinct from the canonical seesaw-based leptogenesis due to the presence of vector-like fermions. We find that leptogenesis is viable for $M_\Sigma \gtrsim 10^{7}$ GeV for a hierarchical spectrum of fermion triplets. However, in the quasi-degenerate regime, resonant enhancement of the $CP$ asymmetry lowers this scale down to $\mathcal{O}({\rm TeV})$, reconciling successful leptogenesis with the originally motivated TeV-scale phenomenology and testability of the model at colliders. 
\end{abstract}

\maketitle

\section{Introduction}
\label{sec:intro}
Neutrino oscillation experiments over the past three decades have firmly established that at least two of the three active neutrino mass eigenstates must have small but nonzero eigenvalues. 
Since neutrinos are predicted to be exactly massless within the Standard Model (SM), these observations provide a concrete laboratory evidence for beyond the Standard Model (BSM) physics.  A particularly elegant explanation of small neutrino masses is provided by the effective dimension-5 Weinberg operator $LLHH/\Lambda$~\cite{Weinberg:1979sa} (where $L$ and $H$ are the SM lepton and Higgs doublets respectively, and $\Lambda$ is the new physics scale),  which generates Majorana neutrino masses after electroweak symmetry breaking (EWSB). In ultraviolet-complete theories, this operator is commonly realized at the tree level through type-I~\cite{Minkowski:1977sc,Mohapatra:1979ia,Yanagida:1980xy,Gell-Mann:1979vob}, type-II~\cite{Konetschny:1977bn,Magg:1980ut,Schechter:1980gr,Cheng:1980qt,Lazarides:1980nt,Mohapatra:1980yp}, or type-III~\cite{Foot:1988aq} seesaw mechanism involving either heavy fermion singlets, $SU(2)_L$ scalar triplets, or $SU(2)_L$ fermion triplets, respectively. 

The Weinberg operator violates lepton number by two units. In the canonical seesaw mechanisms, the scale of lepton number violation (LNV) is typically the same as the mass of the heavy particles responsible for the smallness of neutrino masses. According to 't Hooft naturalness arguments~\cite{tHooft:1979rat}, the natural value of the seesaw scale is $\sim 10^{14}$ GeV, close to the Grand Unified Theory scale. Such high scales associated with canonical dimension-5 seesaw scenarios often make direct experimental tests of LNV rather difficult.\footnote{Lowering the seesaw scale in the minimal scenario requires either tiny Yukawa couplings or approximate conservation of lepton number~\cite{Kersten:2007vk, Ibarra:2010xw}.}  Several mechanisms have been proposed, e.g.~variations of the singlet seesaw with additional fermions~\cite{Mohapatra:1986bd,Akhmedov:1995ip,Akhmedov:1995vm,Malinsky:2005bi}, and radiative seesaw~\cite{Zee:1985id,Babu:1988ki,Pilaftsis:1991ug,Tao:1996vb,Krauss:2002px,Ma:2006km,Dev:2012sg,Cai:2017jrq,Klein:2019iws,Babu:2019mfe}, which can naturally lower the seesaw scale to the TeV range, and hence, enhance its experimental testability~\cite{Deppisch:2015qwa, Cai:2017mow}. 
Another appealing possibility is to generate neutrino masses through higher-dimensional operators~\cite{Babu:2001ex,deGouvea:2007qla,Babu:2009aq,Bonnet:2009ej,Picek:2009is,Liao:2010cc,Kumericki:2011hf,Kumericki:2012bh,
McDonald:2013kca,Wang:2016lve,Cepedello:2017eqf,Anamiati:2018cuq,Dorsner:2019vgf,Giarnetti:2023dcr}. In such scenarios, the tree-level neutrino mass is typically suppressed by factors of the form  $1/\Lambda^{d-4}$, where $d\geq 7$ denotes the operator dimension. As a result of this additional suppression associated with higher-dimensional operators,  the scale of new physics responsible for neutrino mass can naturally be lowered to the TeV range without requiring extremely small Yukawa couplings or fine-tunings. These operators may arise either at the tree level or radiatively at the loop level, and generally necessitate an extended particle spectrum involving new fermions and scalar fields in nontrivial representations of $SU(2)_L$. Since the associated new physics scale can lie around a few TeV, these models offer rich phenomenological prospects, including possible signatures at the Large Hadron Collider (LHC). 

Here, we focus on a particular model realization by Babu, Nandi and Tavartkiladze (BNT) based on a dimension-7 operator involving scalar quadruplets and vector-like fermion triplets~\cite{Babu:2009aq}. The implications of this model for colliders have been studied in Refs.~\cite{Bambhaniya:2013yca, 
Ghosh:2017jbw,Ghosh:2018drw, Pan:2019wwv,Giarnetti:2023dcr, Chakraborty:2025kcl}, for muon $g-2$ in Ref.~\cite{Arbelaez:2020rbq} and the vacuum structure of the model was recently studied in Ref.~\cite{Ashanujjaman:2025rqk}. Note that, in the BNT model the dimension-5 Weinberg operator can also be generated at one loop, and whether the dominant contribution to the neutrino mass comes from the dimension-7 operator or the dimension-5 operator depends on the masses of the new particles~\cite{Bambhaniya:2013yca}.    

 Apart from the origin of neutrino mass, another open problem in particle physics and cosmology is the 
 origin of the observed baryon asymmetry of the Universe (BAU). An elegant framework to address this issue is the mechanism of leptogenesis~\cite{Fukugita:1986hr}, in which a primordial lepton asymmetry generated in the early Universe is partially converted into the observed baryon asymmetry through electroweak sphaleron processes~\cite{Kuzmin:1985mm}. 
What makes this option very attractive is that leptogenesis can occur naturally in seesaw models that inherently incorporate LNV through the Weinberg operator.\footnote{An alternative is Dirac leptogenesis without violating global lepton number~\cite{Dick:1999je, Murayama:2002je}, provided there is a mechanism to source an initial asymmetry between left- and right-handed neutrinos (and their corresponding antiparticles). The electroweak sphalerons partially convert the left-handed lepton asymmetry into a baryon asymmetry, while the tiny Dirac Yukawa couplings prevent the right-handed neutrinos from reaching thermal equilibrium.
}
Conventional leptogenesis is usually realized at very high energy scales $\gtrsim \mathcal{O}(10^9)$ GeV~\cite{Davidson:2002qv, Buchmuller:2002jk}, often associated with canonical seesaw models~\cite{Buchmuller:2004nz,Hambye:2012fh}, making them difficult to test experimentally.\footnote{Including flavor effects can lower the leptogenesis scale to $\sim 10^6$ GeV, but requires certain degree
of cancellation between the tree and one-loop level contribution to the light neutrino mass matrix~\cite{Moffat:2018wke}.}This has motivated growing interest in low-scale leptogenesis frameworks~\cite{Akhmedov:1998qx,Pilaftsis:2003gt,Chun:2017spz}, where the origin of neutrino mass and the BAU can be connected to new physics accessible at present or future experiments. To this extent, it is natural to explore if leptogenesis can be realized in the neutrino mass models of higher-dimensional operators  involving new fermion and scalar ${\bf n}$-tuplets of $SU(2)_L$ (with $\bf{n} \geq 3$), and whether it is viable at the TeV scale.  

To this effect, we study the possibility of thermal leptogenesis in the BNT model. The presence of vector-like fermions can have interesting consequences for leptogenesis, which can be contrasted against the Majorana fermion singlet/triplet in type-I~\cite{Buchmuller:2004nz}/type-III~\cite{Hambye:2003rt} seesaw leptogenesis. Similarly, the presence of a new scalar multiplet in the model can be contrasted with type-II seesaw leptogenesis~\cite{Ma:1998dx,Hambye:2003ka,Hambye:2005tk}.
Building upon the features of leptogenesis with canonical seesaw, we will focus on the decays of heavy vector-like triplets to realize thermal leptogenesis. Unlike the canonical seesaw, the Yukawa couplings in the BNT model are not necessarily suppressed by neutrino masses. The suppression is rather attributed to the small vacuum expectation value (VEV) of the quadruplet $(v_\Phi)$, which in turn is induced from the SM Higgs VEV via a quartic term in the scalar potential. Therefore, the analog of the Davidson-Ibarra (DI) bound~\cite{Davidson:2002qv} for vanilla leptogenesis does not apply here, hence allowing for a lower leptogenesis scale. We find that the scale of leptogenesis can be as low as $3.5\times 10^7$ GeV in the hierarchical case. This can be contrasted against the Majorana fermion singlet (triplet) in type-I (III) leptogenesis, where vanilla leptogenesis occurs only above $10^9$ GeV~\cite{Davidson:2002qv} ($3\times 10^{10}$ GeV~\cite{Hambye:2003rt}). 

In addition to the hierarchical case, we also consider the possibility of low-scale resonant leptogenesis considering a quasi-degenerate spectrum of the vector-like triplet fermions in the BNT model. 
As expected, the scale of leptogenesis in this case can be lowered to the TeV scale. We find a  lower bound of 1.7 TeV on the triplet fermion mass, very close to the theoretical lower bound of 1.6 TeV which applies to both type-II and type-III seesaw  leptogenesis~\cite{Strumia:2008cf}. 

A unique feature of the BNT model is that LNV can originate either from the scalar potential, when the scalar quadruplet carries a lepton number of $-2$, or from the Yukawa interaction if the scalar quadruplet is assigned a zero lepton number.
In the first case, the decay process of the vector-like fermions itself is not lepton number violating,which is similar to the Dirac leptogenesis scenario. 
However, because the vector-like fermion can decay to two different scalars, a net lepton asymmetry can nevertheless be generated if the couplings are different. 
Note that this mechanism differs  from the conventional Dirac leptogenesis because of  the presence of an explicit lepton number violating term in the Lagrangian. Conversely, when the scalar does not carry the lepton number, the decay process itself violates the lepton number, as in the canonical seesaw leptogenesis.

The rest of the article is organized as follows: In Sec.~\ref{sec:model}, we describe the basic features of the BNT model including its scalar and fermion sectors. In Sec.~\ref{sec:numass}, we describe the neutrino mass generation in the BNT model and outline the adopted Yukawa parameterization procedure. We discuss the $CP$ asymmetry generation and the Boltzmann equations relevant for thermal leptogenesis for a hierarchical spectrum of heavy fermion triplets in Sec.~\ref{sec:lepto} and present our results in Sec.~\ref{sec:results}.  
We then discuss the possibility of low-scale leptogenesis and the relevant constraints in Sec.~\ref{sec:resonant}. We finally conclude in Sec.~\ref{sec:conc}. The decay and scattering rates are given in Appendix~\ref{app:rates} and the sphaleron conversion factor specific to this model is calculated in Appendix~\ref{app:Sphaleron}.  

\section{The BNT Model}\label{sec:model}

In addition to the SM particles, the BNT model~\cite{Babu:2009aq} contains a scalar quadruplet $\Phi \sim ({\bf 1},{\bf 4},3/2)$ and $2$ generations of a pair of vector-like fermions $\Sigma_{L,R} \sim ({\bf 1},{\bf 3},1)$, with 
\begin{align}
&\Phi \left({\bf 1},{\bf 4},3/2\right)= \{\; \Phi^{+++}, \Phi^{++}, \Phi^{+}, \Phi^{0}\;  \}, \\
&\Sigma_{L,R} ({\bf 1},{\bf 3},1)= \{ \; \Sigma_{L,R}^{++}, \Sigma_{L,R}^{+}, \Sigma_{L,R}^{0} \; \}. 
\label{eq:sigma}
\end{align}
The particles and their charge assignments are given in Table~\ref{tab:part}. We also show the global lepton number (${\sf L}$) assignments for the fields. 

Note that $\Phi = \Phi_{ijk}$ ($i,j,k=1,2$) is a totally symmetric tensor of rank 3, whose components are given by 
\begin{align}
    &\Phi_{111} = \Phi^{+++}\,,\quad \Phi_{112,121,211} = \frac{\Phi^{++}}{\sqrt{3}}\,,\nonumber\\
    &\Phi_{122,212,221} = \frac{\Phi^{+}}{\sqrt{3}}\,,\quad \Phi_{222} = \Phi^0\,.
\end{align}
Similarly, $\Sigma = \Sigma_{ij}$ is a  totally symmetric tensor of rank 2, and its components are
\begin{align}
&\Sigma_{11} = \Sigma^{++},\, \quad  \Sigma_{22} = \Sigma^0,\, \quad 
\Sigma_{12,21}= \frac{\Sigma^+}{\sqrt{2}}\,.
\end{align}
The relevant Lagrangian for the model reads as
\begin{align}
&\mathcal{L} \supset\,  
\mathcal{L}_{\rm kin}+\mathcal{L}_{\rm yuk}  - \mathcal{V}(H,\Phi)\,,
\label{eq:Lagrangian}
\end{align}
where, suppressing the $SU(2)$ indices, the kinetic and Yukawa terms are respectively given by 
\begin{align}
\mathcal{L}_{\rm kin} = & \  (D^\mu H)^\dagger(D_\mu H) + 
 (D^\mu \Phi)^\dagger(D_\mu \Phi) \nn \\
& \quad +\overline{\Sigma}_{L,R\;j} \;i\gamma^{\mu} D_{\mu} \Sigma_{L,R\;j} \,,\label{eq:kin}
\\
-\mathcal{L}_{\rm yuk} =& \ 
M_{\Sigma\,ij}
 \overline{ \Sigma_L }_{i} \Sigma_{Rj} + {Y_H}_{\alpha i}\; \overline{(L_{\alpha})^c} \,\epsilon \,\Sigma_{L\;i}  H^\ast \nonumber\\
 &\quad\quad+ {Y_{\Phi}}_{\alpha i}\; \overline{L}_{\alpha} \,\epsilon\, \Phi^{\ast} \Sigma_{R\;i} \;+ \hc \,,\label{eq:yuk}
\end{align}
where $\epsilon$ is the $SU(2)$ antisymmetric tensor,  $D_\mu \equiv \partial_\mu - i g {\cal T}_a W_\mu^a - i g'(Y/2)\, B_\mu$ is the covariant derivative leading to gauge interactions of the BSM multiplets with the $SU(2)_L$ and $U(1)_Y$ gauge bosons $W_\mu$ and $B_\mu$ respectively, and ${\cal T}_a$ generically stand for the $SU(2)$ generators in the isospin 1/2 (doublet), 1 (triplet) and 3/2 (quadruplet) representations for $H$, $\Sigma$ and $\Phi$ fields,  respectively.  
\begin{table}[!t]
\centering
\begin{tabular}{c c c}
\toprule
Field & $SU(3)_c\times SU(2)_L \times U(1)_Y$ & ${\sf L}$ \\
\midrule
$L$ & $\left( {\bf 1}, {\bf 2} , -1/2 \right)$ & $+1$ \\
$H$ & $\left( {\bf 1}, {\bf 2}, 1/2 \right)$ & $0$ \\
\hline
$\Sigma$ & $\left({\bf 1}, {\bf 3}, 1 \right) $ & $-1/+1$\\
$\Phi$ & $\left({\bf 1}, {\bf 4}, 3/2 \right)$ & $-2/0$ \\
\bottomrule
\end{tabular}
\caption{The particle content of the BNT model and their charge assignments under the SM gauge group $SU(3)_C \times SU(2)_L \times U(1)_Y$. We also show the lepton number  assignments: $\Sigma$ can be assigned either ${\sf L} = -1$ or $+1$, and similarly $\Phi$ can carry ${\sf L}$ = $-2$ or 0.} 
\label{tab:part}
\end{table}

The Yukawa couplings $Y_{H}$ and $Y_{\Phi}$ are $3 \times 2$ complex matrices with $\alpha , \beta =e, \mu, \tau$, and $i,j=1,2$, each characterized by 12 real parameters:
\begin{align}\label{eq:yukmat}
Y_{\Phi}= \begin{pmatrix}
Y^{\Phi}_{e 1}&Y^{\Phi}_{e 2}\\
Y^{\Phi}_{\mu 1}&Y^{\Phi}_{\mu 2}\\
Y^{\Phi}_{\tau 1}&Y^{\Phi}_{\tau 2}
\end{pmatrix},\, \qquad 
Y_{H}= \begin{pmatrix}
Y^{H}_{e1}&Y^{H}_{e2} \\
Y^{H}_{\mu 1}&Y^{H}_{\mu2}\\
Y^{H}_{\tau 1}&Y^{H}_{\tau2}\\
\end{pmatrix}\,.
\end{align}
The Dirac mass term for the triplet, $M_{\Sigma}$, is taken to be real and diagonal, without loss of generality, so $M_\Sigma = \text{Diag}(M_{\Sigma_1},M_{\Sigma_2})$.

The scalar potential of the model can be written as
\begin{align}
{\mathcal V} \left(H,\Phi \right) &= -\mu_H^2  \left(H^\dagger H \right) + \lambda_1 \left(H^\dagger H \right)^2 \nonumber\\
&+ \mu_{\Phi}^2\left(\Phi^\dagger \Phi \right)+ \lambda_2\left(\Phi^\dagger \Phi \right)^2   \nonumber \\ 
&+\lambda_3(H^\dag H)(\Phi^\dag \Phi)+\lambda_4(H^\dag \tau_a H)(\Phi^\dag T_a \Phi) \nonumber\\
&+ \left(\lambda_5  H^3\, \Phi^{\ast} +\hc\right) \,,\label{eq:potential}
\end{align}
where we take all the parameters to be real. Here $\tau_a=\sigma_a/2$ (with $\sigma_a$ being the standard Pauli matrices) and $T_a$ stand for the $SU(2)$ generators in the isospin 1/2 and 3/2 representations, respectively. 

Upon integrating out the $\Sigma$ field in Eq.~\eqref{eq:yuk}, a Weinberg-like dimension-5 operator is obtained:
\begin{equation}
\label{eq:effop5}
    \mathcal{L}_5 = -\frac{(Y_\Phi Y_H^\dagger + Y_H^\ast Y_\Phi^T)\,\overline{L}L^c\,H \Phi}{M_\Sigma}+\hc\,,
\end{equation}
where we have suppressed the generation indices. It can be seen that once the scalars take a VEV, a Majorana mass is generated for the neutrinos, $m_\nu \propto v_\Phi v_H$. 

Before discussing the scalar and fermion sectors of the model in detail, we briefly comment on the source of LNV in the model -- a crucial ingredient for leptogenesis. Similar to the case of type-II seesaw, LNV in the model can be attributed to either the Yukawa terms or to the scalar potential, depending on the ${\sf L}$ charge assigned to $\Phi$. First, let us consider the case where $\Phi$ carries a charge ${\sf L}=-2$ units while $\Sigma_{L,R}$ is assigned $-1$. In this case, it can be seen that the $\lambda_5$ term in Eq.~\eqref{eq:potential} softly breaks ${\sf L}$ number, while $\mathcal{L}_{\rm yuk}$ preserves it~\cite{Babu:2009aq}. In the limit, $\lambda_5 \rightarrow 0$, the symmetry of the scalar potential is restored, therefore, its value can be naturally small in the 't Hooft sense~\cite{tHooft:1979rat}. 
Alternatively, if $\Phi$ has ${\sf L}$ = $0$ charge, and $\Sigma_{L,R}$ carry a unit lepton number, the either $Y_H$ or $Y_{\Phi}$ term in $\mathcal{L}_{\text{yuk}}$ violates ${\sf L}$ number by 2 units, and must be suppressed~\cite{Bambhaniya:2013yca}. Since lepton number is violated by two units (owing to the Weinberg operator) in both the scenarios, one can adopt either assignment, as shown in Table \ref{tab:part}, to realize leptogenesis. We discuss the implications later.

\subsection{Scalar Sector}

After EWSB, the Higgs VEV, $v_H \simeq 246.2$ GeV induces a non-zero VEV for the neutral component of $\Phi$ via the $\lambda_5$ term:
\begin{equation}
    v_\Phi = -\frac{\lambda_5 v_H^3}{2M_{\Phi^0}^2}\simeq 0.74~{\rm GeV}\, \left(\frac{\lambda_5}{0.1}\right)\left(\frac{1~{\rm TeV}}{M_{\Phi^0}}\right)^2\,.
    \label{eq:vphi}
\end{equation}

Since a scalar quadruplet obtaining a VEV breaks the custodial symmetry, this induced VEV is further constrained by modifications to the $\rho$ parameter: $\rho \simeq (1-6v_\Phi^2/v_H^2)$. Using the experimental value of the $\rho$ parameter~\cite{ParticleDataGroup:2024cfk}, the VEV is constrained to be much smaller than the SM Higgs VEV: 
\begin{equation}\label{eq:rhovev}
    v_\Phi < 2.6~\text{GeV}\,,
\end{equation}
which implies that $\lambda_5 \lesssim 0.35$ for $M_\Phi \sim \mathcal{O}(1)$ TeV. 

Further constraints on the scalar quadruplet can be placed from its production and decays at the LHC. The effective operator in Eq.~\eqref{eq:effop5} leads to an interesting phenomenological signal, namely the decays of the doubly charged component $\Phi^{\pm\pm}$ into a pair of same-sign leptons ($\Phi^{\pm\pm} \rightarrow \ell^{\pm} \ell^{\pm} $), after EWSB. The strength of this interaction is proportional to the active neutrino mass matrix, analogous to the case of Type-II seesaw, and goes like $v_\Phi^{-1}$. These decays to same-sign dileptons compete against decays to gauge bosons $\Phi^{\pm\pm}\rightarrow W^\pm W^\pm$, which are proportional to $v_\Phi$. Therefore, depending on the value of $v_\Phi$, the dominant decay channel is either leptonic $(v_\Phi \lesssim \mathcal{O}(100~\text{keV}))$ or bosonic $(v_\Phi \gtrsim \mathcal{O}(100~\text{keV}))$.\footnote{There are also cascade decays of $\Phi^{\pm\pm}$ to other components of the quadruplet (and vice versa), and further sequential decays to leptons or gauge bosons. See Refs.~\cite{Ghosh:2018drw,Giarnetti:2023dcr} for further discussion.} Current searches for the decays of doubly-charged scalars to multi-lepton final states~\cite{CMS:2017pet,ATLAS:2022pbd} or to a pair of gauge bosons~\cite{CMS:2017fhs,ATLAS:2021jol} at the LHC lead to the following bounds~\cite{Giarnetti:2023dcr}:
\begin{align}
    M_{\Phi^{\pm\pm}} > \begin{cases}
        860~\text{GeV},\quad\text{Br}( \ell^{\pm} \ell^{\pm}) = 100\%\, ,\\
        260~\text{GeV},\quad\text{Br}( W^\pm W^\pm) = 100\% \, .
    \end{cases}
\end{align}
Since the mass splitting between the different components of the quadruplet is restricted to be $\Delta M\lesssim 50$ GeV from electroweak precision data~\cite{Ghosh:2018drw}, the lower limits mentioned above should generically apply to all components of $\Phi$.  
To be safe from these constraints in general, we consider $M_\Phi \gtrsim 1$ TeV in our numerical analysis.

\subsection{Fermion Sector}

The model contains three neutral left-handed fermion fields $\nu_L, \Sigma^0_L, \Sigma^{0,c}_R$, with the following mass matrix:
\begin{align}\label{neutral}
-\mathcal{L}_{\nu \Sigma^0}  = &  \frac{1}{2} 
\begin{pmatrix}
\overline{\nu_L^c} &   \overline{\Sigma^{0,c}_L} & \overline{{\Sigma^{0}_R}}
\end{pmatrix}
\begin{pmatrix}
0 &\frac{v_H}{\sqrt{2}} Y_H & \frac{v_{\Phi}}{\sqrt{2}} Y^*_{\Phi}\\
\frac{v_H}{\sqrt{2}} Y^T_H &0 & M_{\Sigma}\\
\frac{v_{\Phi}}{\sqrt{2}} Y^\dag_{\Phi}&M_{\Sigma}&0
\end{pmatrix}
\begin{pmatrix}
\nu_L\\  {\Sigma^{0}_L} \\ \Sigma^{0,c}_R
\end{pmatrix}\, \nn \\
& \qquad +\hc
\end{align}
If only one generation of vector-like fermion is introduced, this is a $5\times 5$ matrix with rank 4, leading to two active neutrino masses, which can be contrasted with the case of Type-III seesaw where at least two generations of fermion triplets are required for two non-zero neutrino masses. While a single $\Sigma$ generation is sufficient to explain two non-zero active neutrino masses, the generation of $CP$ asymmetry via $\Sigma$ decay requires at least two generations of $\Sigma$, as we discuss below. Therefore, for our case, the resulting neutrino mass matrix is $7\times 7$ which has rank 7 and leads to 3 non-zero light neutrino masses.

Similarly, the singly charged component of $\Sigma$ mixes with the SM charged leptons. From Eq.~\eqref{eq:sigma}, we see that $\Sigma_L^+$ is positively-charged with left-chirality and $\Sigma_R^+$ is positively-charged with right-chirality. Similarly, $(\Sigma_R^+)^c$ is negatively-charged with left chirality, and  $(\Sigma_L^+)^c$ is negatively-charged with right chirality. Thus the Dirac mass matrix in the charged fermion sector connecting the left- and right-chiral fields is given by 
\begin{align}\label{charged}
-\mathcal{L}_{\ell \Sigma} = &  
\begin{pmatrix}
\overline{\ell^-_L} & \overline{(\Sigma^+_R)^c}
\end{pmatrix}
\begin{pmatrix}
m_\ell  & {v_H} Y^*_H/2\\
0&M_{\Sigma}
\end{pmatrix}
\begin{pmatrix}
\ell^-_R\\  (\Sigma^+_L)^c
\end{pmatrix}\nn \\
& \qquad +\hc \,
\end{align}
The mass eigenstates can be found by diagonalizing the above mass matrix. 

Both CMS and ATLAS collaborations at the LHC have put stringent limits on the mass of the triplet fermions by analyzing final states including production of up to four light leptons, and up to three hadronically decaying $\tau$ leptons~\cite{ATLAS:2022yhd, CMS:2022nty}, leading to $M_\Sigma > 1065$ GeV. Therefore, the mixing between the SM leptons and triplets is expected to be suppressed by $m_\ell/M_{\Sigma}$, and the weak states almost coincide with the mass eigenstates for the charged leptons.

Since we are interested in the decays of $\Sigma$ for leptogenesis, we work with the following hierarchy: $M_{\Sigma}> M_\Phi \gg M_{H}$, which leads to the following decay channels: $\Sigma \r L\Phi\,,\bar{L}H$ and $\bar{\Sigma} \r \bar{L}\Phi^{\ast}\,,LH^{\ast}$. The total decay widths and branching ratios can be found in Appendix~\ref{app:rates}.

\section{Neutrino Mass}\label{sec:numass}

\begin{figure}
\includegraphics[width=0.49\textwidth]{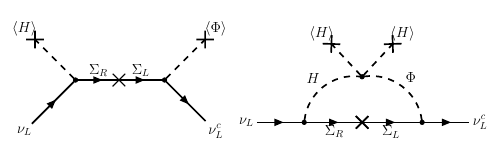}
 \caption{Feynman diagrams for the tree and one-loop contributions to neutrino mass in the BNT model.}
     \label{fig:numassdiagram}
\end{figure}

The effective operator in Eq.~\eqref{eq:effop5} generates tree-level masses for the light neutrinos after EWSB as follows: 
\begin{align}
m_{\nu}^{\text{tree}}=& -v_H v_\Phi\left(
Y_{\Phi} M^{-1}_{\Sigma} Y_H^{\dagger}  +  
Y^\ast_H M^{-1}_{\Sigma} Y^T_{\Phi}
\right) \nonumber\\
=& - \frac{\lambda_5 v_H^4}{2M^2_{\Phi^0}} 
\left(
Y_{\Phi} M^{-1}_{\Sigma} Y_H^{\dagger}  +  
Y^\ast_H M^{-1}_{\Sigma} Y^T_{\Phi}
\right)\,,
\end{align}
where the second line is obtained after integrating $\Phi$ out [cf.~Eq.~\eqref{eq:vphi}], resulting in a dimension-7 neutrino mass; see Fig.~\ref{fig:numassdiagram} left panel. Notice the additional suppression from $\lambda_5$ and $M_\Phi^2$, compared to type-I seesaw where $m_\nu \simeq -v_H^2 Y M_N^{-1} Y^\dagger$. 

Moreover, the same $\lambda_5$ term responsible for the induced VEV of $\Phi$ also allows us to write the dimension-5 Weinberg operator at one-loop level, see Fig.~\ref{fig:numassdiagram} right panel, with the loop contribution given by~\cite{Giarnetti:2023osf}
\begin{align}
m_{\nu}^{\text{loop}}=\frac{\lambda_5 v_H^2}{4\pi^2}\frac{\left(
Y_{\Phi} Y_H^{\dagger}  +  
Y^\ast_H Y^T_{\Phi}
\right)M_\Sigma}{M_\Phi^2-M_H^2}\,f(M_\Phi, M_H, M_\Sigma)\,,
\end{align}
where the loop function is given by
\begin{align}
    f(x,y,z) = \frac{x^2}{x^2-z^2}\ln\left(\frac{x^2}{z^2}\right)-\frac{y^2}{y^2-z^2}\ln\left(\frac{y^2}{z^2}\right)\,.
\end{align}
The total active neutrino mass is given by the sum 
\begin{align}
    m_\nu = m_\nu^{\rm tree} +  m_\nu^{\rm loop}\,.
\end{align}

\begin{figure}[!t]
    \centering
    \includegraphics[width=0.95\linewidth]{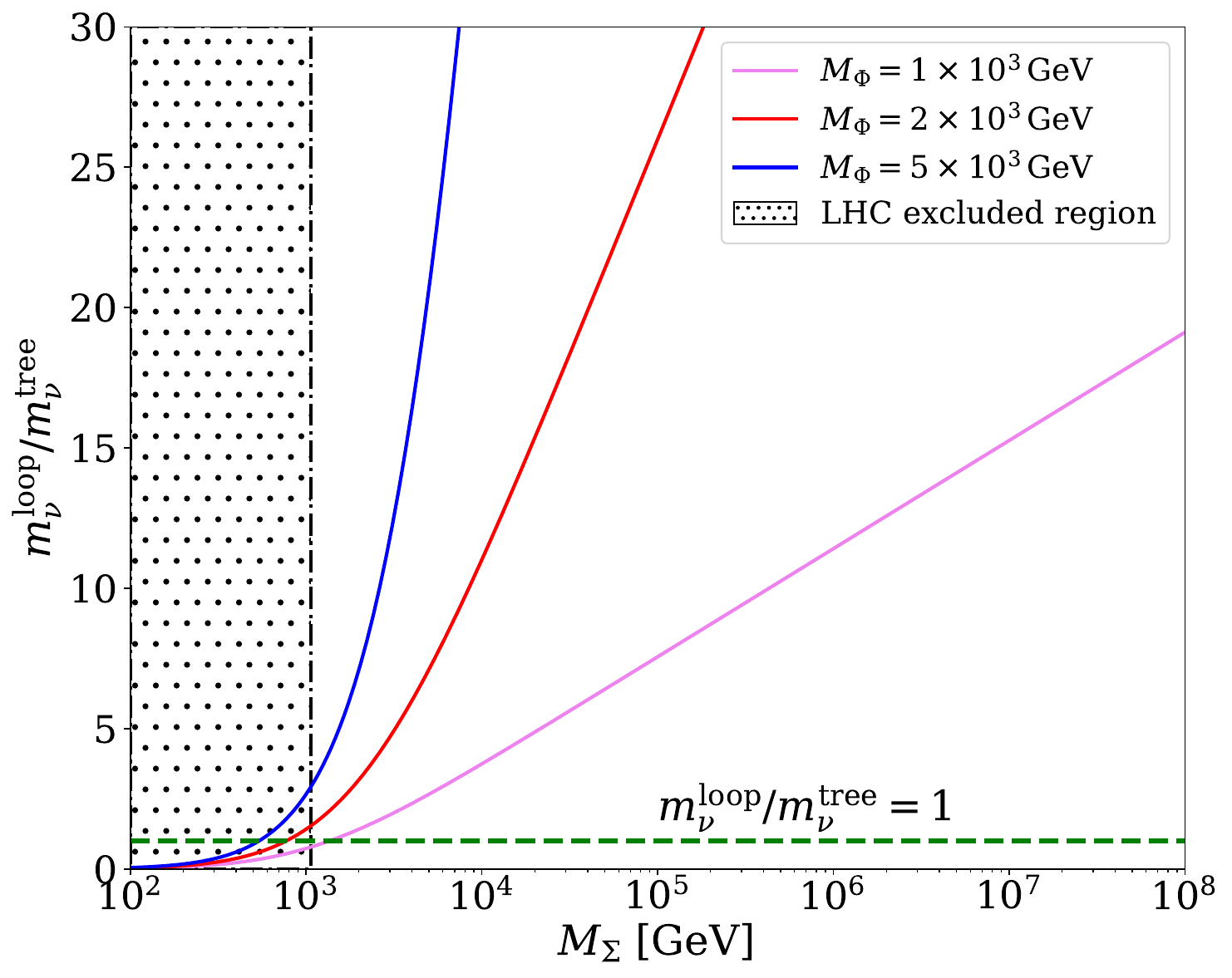}
    \caption{Comparison of the loop (dimension-5) and tree (dimension-7) contributions to neutrino mass in the BNT model  with respect to $M_\Sigma$ for different values of $M_\Phi$. The green dashed line represents $m_\nu^{\rm loop}/m_\nu^{\rm tree} =1$, whereas the vertical shaded region is excluded from LHC searches for triplet fermions. }
    \label{fig:numass}
\end{figure}

It was shown in Ref.~\cite{Bambhaniya:2013yca}  that the loop contribution can dominate over the tree-level one in a wide region of the parameter space. In Fig.~\ref{fig:numass}, we compare the loop and tree contributions to neutrino mass with respect to $M_\Sigma$ for fixed values of $M_\Phi$. It can be seen that for the allowed range of $\Sigma$ masses above the TeV scale, the dimension-5 operator-induced loop contribution is dominant over the dimension-7 operator-induced tree contribution and the loop dominance increases with the higher values of $M_{\Phi}$. 

Next, we discuss the parameterization of the Yukawa matrices of Eq.~\eqref{eq:yukmat} to satisfy the constraints from the low-energy neutrino sector, as the Yukawa values play an important role in determining the $CP$ asymmetry produced in $\Sigma$ decays.

Owing to two sets of distinct Yukawa matrices $Y_\Phi$ and $Y_H$ in the BNT model, as opposed to a single Yukawa matrix in canonical seesaw, we adopt the Generalized Casas-Ibarra (GCI) parameterization approach outlined in Ref.~\cite{Herrero-Garcia:2025aox}. The neutrino masses in the model following the hierarchy $M_\Sigma > M_\Phi \gg M_H$ can be written as
\begin{align}
    m_\nu = \mathcal{Y}^T \mathcal{M} \mathcal{Y}\,,
\end{align}
where $\mathcal{Y} = (Y_\Phi^T, Y_H^\dag)^T$ is a $4\times 3$ complex matrix [cf.~Eq.~\eqref{eq:yukmat}], and 
\begin{align}
    \mathcal{M} \simeq \frac{\lambda_5 v_H^2}{4\pi^2\, M_\Phi^2}\begin{pmatrix}
        0 & 0 & \frac{f'_1}{M_{\Sigma_1}}& 0\\
        0 & 0 & 0 & \frac{f'_2}{M_{\Sigma_2}}\\
        \frac{f'_1}{M_{\Sigma_1}} & 0 & 0 & 0\\
        0 & \frac{f'_2}{M_{\Sigma_2}} & 0 & 0
    \end{pmatrix}\,,
\end{align}

where 
\begin{align}
    f'_i = M_\Phi^2 \ln(M_{\Sigma_i}^2/M_\Phi^2)-M_H^2 \ln(M_{\Sigma_i}^2/M_H^2)\,,
\end{align}
where we have neglected the tree-level contribution to neutrino masses as they are sub-dominant (cf.~Fig.~\ref{fig:numass}). The Yukawa couplings can then be written as~\cite{Casas:2001sr, Herrero-Garcia:2025aox}
\begin{equation}
    \mathcal{Y} = V^\dagger D_\mathcal{M}^{-1/2} \mathcal{R} D_{m}^{1/2} U^\dagger\,,
\end{equation}
where $V$ is the unitary matrix that diagonalizes $\mathcal{M}$ such that $\mathcal{M} = V^T D_\mathcal{M} V$, and is given by
\begin{align}
    V = \frac{1}{\sqrt{2}}\begin{pmatrix}
        i & 0 & -i & 0\\
        0 & i & 0 & -i\\
        1 & 0 & 1 & 0\\
        0 & 1 & 0 & 1
    \end{pmatrix}\,,
\end{align}
$U$ is the PMNS matrix diagonalizing the light neutrino mass matrix: $D_m = U^T m_\nu U$,  and $\mathcal{R}$ is a complex semi-orthogonal $4 \times 3$ matrix
\begin{eqnarray}
    {\cal R} & =& {\cal R}'_{4 \times 4} \cdot \begin{pmatrix}
        {\mathbb{I}_{3\times3}}\\
        {0_{1 \times 3}}
    \end{pmatrix}\, ,
\end{eqnarray}
where ${\cal R}' \in SO(4)$ and is characterized by 6 complex angles $\theta_{ij}$.\footnote{Some of these angles can be set to 0, for example, depending upon any imposed flavor symmetries or whether the lightest active neutrino remains massless. For instance, we find that fixing $\theta_{23},\theta_{24} = 0$ enhances the $CP$ asymmetry.}

\section{Leptogenesis}\label{sec:lepto}
\begin{figure*}
\includegraphics[width=0.9\textwidth]{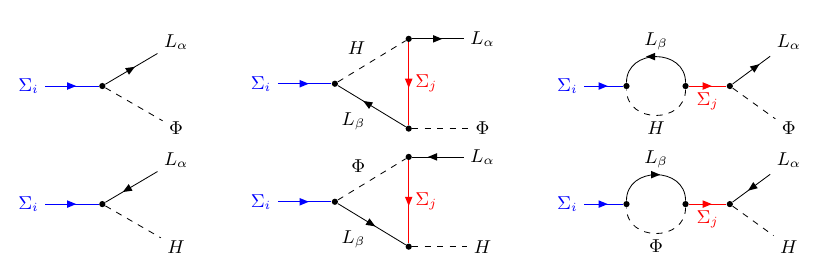}
 \caption{Tree and one-loop diagrams for the decays of $\Sigma$ to $L\Phi$ and $\bar{L}H$ that produce the $CP$ asymmetry in the BNT model. }
     \label{fig:lepto}
\end{figure*}

With the assumed hierarchy $M_{\Sigma}> M_\Phi \gg M_{H}$, the heavy fermion triplets can decay out-of-equilibrium to the SM leptons along with either the SM Higgs $H$ or the quadruplet scalar $\Phi$ (and their conjugates) via the complex Yukawa interactions $Y_H$ and $Y_\Phi$ respectively. We define the yields
\begin{align}
    Y_i \equiv \frac{n_i}{s},\quad Y_{\Delta i} \equiv Y_i - Y_{\bar{i}}\,,
\end{align}
where $n_i$ are the corresponding number densities, and $s = (2\pi^2/45)g_\ast T^3$ is the co-moving entropy density of the Universe, with $g_*$ being the effective relativistic degrees of freedom and $T$ being the temperature.

If $\Phi$ carries a lepton number, these decays themselves do not violate lepton number. A non-vanishing $CP$ asymmetry ($\varepsilon_\Sigma$) is nevertheless produced in their decays due to the interference of the tree and one-loop level diagrams shown in Fig.~\ref{fig:lepto}, leading to a net asymmetry in the lepton doublets ($Y_{\Delta L}$), as well as in the scalars $(Y_{\Delta \Phi},Y_{\Delta H})$. This in turn induces an asymmetry among the triplets ($Y_{\Delta \Sigma}$). Because $M_\Phi \gg M_H$, the lepton number stored in the $\Phi$ sector is eventually washed out later via the $\Phi \r HHH$ decay sourced by the $\lambda_5 H^3 \Phi^\ast$ term in the scalar potential. On the other hand, when $\Phi$ does not carry a lepton number, either the decay channel $\Sigma \r \bar{L}H$ (and $\Sigma \r LH^{\ast}$) or $\Sigma \r L\Phi$ (and $\Sigma \r \bar{L} \Phi^{\ast}$)  violates LN, while the decays to $\Phi$ conserve it. In either case, the $SU(2)_L$ sphalerons reprocess the asymmetry in the SM leptons to a baryon asymmetry
\begin{align}
    Y_{\Delta B} = 3\, c_{\text{sph}} Y_{\Delta L}\,,
\end{align}
where $c_{\text{sph}} = 0.461$ is the sphaleron conversion factor derived in Appendix~\ref{app:Sphaleron}, and the factor of 3 comes from the three components of $\Sigma$. The per component lepton asymmetry can be expressed as
\begin{align}
    Y_{\Delta L} \simeq \vs\, \eta\, Y_\Sigma^{\eq}(T \gg M_\Sigma)=\vs \eta \frac{135\zeta(3)}{2\pi^4 g_\ast}\,,
\end{align}
in terms of the $CP$ asymmetry $\varepsilon_{\Sigma}$ and efficiency parameter $\eta$ such that $0 < \varepsilon_{\Sigma}, \eta \lesssim 1$. The $\eta$ parameter incorporates the effects of washouts, and must be determined via the solution of the Boltzmann equations, with $\eta=1$ implying maximal efficiency (negligible washouts). The equilibrium yield $Y_\Sigma^{\eq} = n_\Sigma^{\eq}/s$ is the total number of triplets ($n_\Sigma+n_{\bar{\Sigma}}$) per unit entropy, and $g_\ast = 114.75$ is the total number of relativistic degrees of freedom at high temperatures, obtained after accounting for the $\Phi$ multiplets. In order to match the present day value of $Y_{\Delta B}^0 = (8.58 \pm 0.17) \times 10^{-11}$~\cite{ParticleDataGroup:2024cfk}, we would thus require $Y_{\Delta L}^0 = (1.92 \pm 0.04) \times 10^{-10}$.

Similar to the case of Type-III seesaw, these fermion triplets have gauge interactions that are in equilibrium for $T \gg M_\Sigma$, but get doubly  Boltzmann suppressed at $T < M_\Sigma$. Unless the Yukawa interactions are strong enough ($Y_{H,\Phi} \gtrsim \mathcal{O}(0.01)$), these gauge annihilations typically dominate and deplete the $\Sigma$ abundance. Subsequently, the decays of the remnant triplet population are not enough to generate the lepton asymmetry required to match the observed BAU. 

As the Yukawa couplings in the BNT model can be much larger compared to the canonical seesaws, one can easily be in the Yukawa-dominated regime to reproduce the BAU for $M_\Sigma \lesssim 10^{10}$ GeV -- below the DI-equivalent bound for Type-III seesaw leptogenesis. The consequence of having large Yukawas is analogous to Type-III seesaw leptogenesis with a second Higgs doublet with $v_{H_2} \ll v_\Phi$, where the lower bound on the Majorana triplet was found to be $\sim \mathcal{O}(10^7)$ GeV for a hierarchical spectrum of triplet masses~\cite{Vatsyayan:2022rth}.  

\subsection{$CP$ Asymmetry}

For a hierarchical spectrum, $M_{\Sigma_2} \gg M_{\Sigma_1}$, we consider only the decays of the lightest fermion triplet $\Sigma_1$, as the asymmetry generated in $\Sigma_2$ decays is washed out by $\Sigma_1$ interactions. Therefore, we drop the subscript to  simplify the notation: $\Sigma_1 \equiv \Sigma$.
The $CP$ asymmetry is generated via the interference of the tree-level and loop-level diagrams shown in Fig.~\ref{fig:lepto}. Notice that the scalar in the external state is different from the one in the internal loop due to the vector-like nature of $\Sigma$. 

The $CP$ asymmetry for the two different decay channels is defined as
\begin{align}
    \varepsilon_H &= \frac{ \sum_{\alpha } \left[ \Gamma \left(\Sigma \rightarrow \bar{L} \, H \right) - \Gamma \left(\bar{\Sigma} \rightarrow L \, H^{\ast}\right) \right]}{2\,\Gamma_\Sigma }\,,\nonumber\\
    \varepsilon_\Phi &= \frac{ \sum_{\alpha } \left[ \Gamma \left(\Sigma \rightarrow L \, \Phi \right) - \Gamma \left(\bar{\Sigma} \rightarrow \bar{L} \, \Phi^{\ast}\right) \right]}{2\,\Gamma_\Sigma }\,,
\end{align}
where $\Gamma_\Sigma = \Gamma_{\bar{\Sigma}}$ is the total decay width of $\Sigma,\bar{\Sigma}$ (hence the factor of 2 in the denominator). An explicit calculation leads to the following result:
\begin{align}\label{eq:cpasym}
    \varepsilon_H &= \frac{1}{2\pi}\frac{\text{Im}[\mathbf{Y}]{f}_{s} \left(z_\Sigma,z_{\Phi}\right) + 2\text{Im}[\mathbf{Y}']{f}_{v} \left(z_\Sigma,z_{\Phi}\right)}{3\left( Y_{H}^{\dagger} \, Y_H \right)_{11} + 4\left( Y_{\Phi}^{\dagger} \, Y_\Phi \right)_{11}}\,,\n\\
    \varepsilon_\Phi &= \frac{1}{2\pi}\frac{\text{Im}[\mathbf{Y}^\dagger]{f}_{s} \left(z_\Sigma,z_{H}\right) + 2\text{Im}[\mathbf{Y}'^\dagger]{f}_{v} \left(z_\Sigma,z_{H}\right)}{3\left( Y_{H}^{\dagger} \, Y_H \right)_{11} + 4\left( Y_{\Phi}^{\dagger} \, Y_\Phi \right)_{11}}\,,
    \end{align}
where $z_\Sigma = M_{\Sigma_2}^2/M_{\Sigma_1}^2\,,z_\Phi = M_\Phi^2/M_{\Sigma_1}^2$, $z_H=M_H^2/M_{\Sigma_1}^2$, and 
\begin{align}
    \mathbf{Y}&= (Y_{H}^{\dagger} Y_H)_{12} (Y_{\Phi}^{\dagger} Y_\Phi)_{21}\,,\n\\
    \mathbf{Y}'&=(Y_{H}^{\dagger}Y_\Phi^{\ast})_{12} (Y_H^T Y_{\Phi})_{21}\,.
\end{align}
The loop functions $f_{s,v}$ corresponding to the contribution from the self-energy and vertex correction, respectively, are given by 
\begin{align}
    f_v(x,y) &= \sqrt{x}\left[1-y-(1+x-y)\ln\left(1+\frac{1}{x}-\frac{y}{x}\right)\right]\,,\n \\
    f_s(x,y) &= \frac{\sqrt{x}}{1-x}(1-y)^2\,. \label{eq:loopfn}
\end{align}
It is straightforward to see that $\varepsilon_\Phi = -\varepsilon_H$ in the limit $z_\Phi, z_H \to 0$, a consequence of CPT and unitarity \cite{Kolb:1979qa,Berbig:2022pye}. The total $CP$ asymmetry generated in $\Sigma$ decays thus becomes
\begin{equation}
    \varepsilon_\Sigma = \varepsilon_\Phi - \varepsilon_H = 2 \varepsilon_\Phi\,.
\end{equation}
The form of the products $\mathbf{Y}$ and $\mathbf{Y}'$ suggest that in the presence of a large hierarchy among the Yukawas, i.e., ${Y_H} \gg {Y_\Phi}$ or ${Y_H} \ll {Y_\Phi}$, the $CP$ asymmetry gets suppressed\footnote{This is the case when $\Phi$ does not carry ${\sf L}$ and the lepton number is violated in either of the Yukawa couplings.}. Instead, it becomes maximal when ${Y_\phi} \simeq {Y_H}$, implying that $\text{Br}_\Phi \simeq 0.6$ and $\text{Br}_H \simeq 0.4$, see Appendix~\ref{app:rates}.\footnote{ If ${Y_{\Phi\alpha i}} = {Y_{H\alpha i}}$, the imaginary part of $\mathbf{Y}$ and $\mathbf{Y}'$ is zero, and the $CP$ asymmetry vanishes.}

Now we discuss whether there exists a bound on the maximal $CP$ asymmetry in the model, analogous to the DI bound for Type-I seesaw~\cite{Davidson:2002qv}. Unlike the case of Type-I or Type-III seesaw, the $CP$ asymmetry and neutrino masses involve different contractions of Yukawa matrices and a standard DI bound can not be defined.\footnote{Similarly, we cannot relate the $\Gamma_\Sigma$ to $m_\nu$ for the analysis as opposed to the case of Type-II and Type-III seesaw, see Ref.~\cite{Hambye:2012fh} for more details.} As the factors of $\lambda_5$ and loop suppression $f'$ that suppress the neutrino masses allowing for $\mathcal{O}(1)$ Yukawas do not appear in the expressions for $\vs$, the $CP$ asymmetry can be as large as $\mathcal{O}(1)$. However, such large Yukawas would also enhance $\Delta L = 2$ scatterings such as $L\Phi \leftrightarrow \bar{L}H$ mediated by $\Sigma_i$ that can washout the produced lepton asymmetry. These scattering processes scale as
\begin{align}
    \Gamma_{\Delta L=2} \simeq \frac{\abs{Y_H}^2\abs{Y_\Phi}^2}{32\pi M_\Sigma^2} T^3\,.
\end{align}
Demanding that these scatterings are out-of-equilibrium, ${\Gamma_{\Delta L =2}}/{H}<1$ at $T \sim M_{\Sigma_1}$ gives us 
\begin{align}
    \abs{y} \lesssim 0.1 \left(\frac{M_{\Sigma_1}}{10^{12}~\text{GeV}}\right)^{1/4},
\end{align}
where $\abs{Y_\Phi} \sim \abs{Y_{H}} \sim \abs{y}$, and  $H=1.66 \sqrt{g_\ast}\,T^2/M_{\text{Pl}}$ is the Hubble rate with 
$M_{\text{Pl}}=1.22 \times 10^{19}$ GeV being the Planck scale. 

\subsection{Boltzmann Equations}
\begin{figure*}[!t]
\centering
\includegraphics[width=0.49\linewidth]{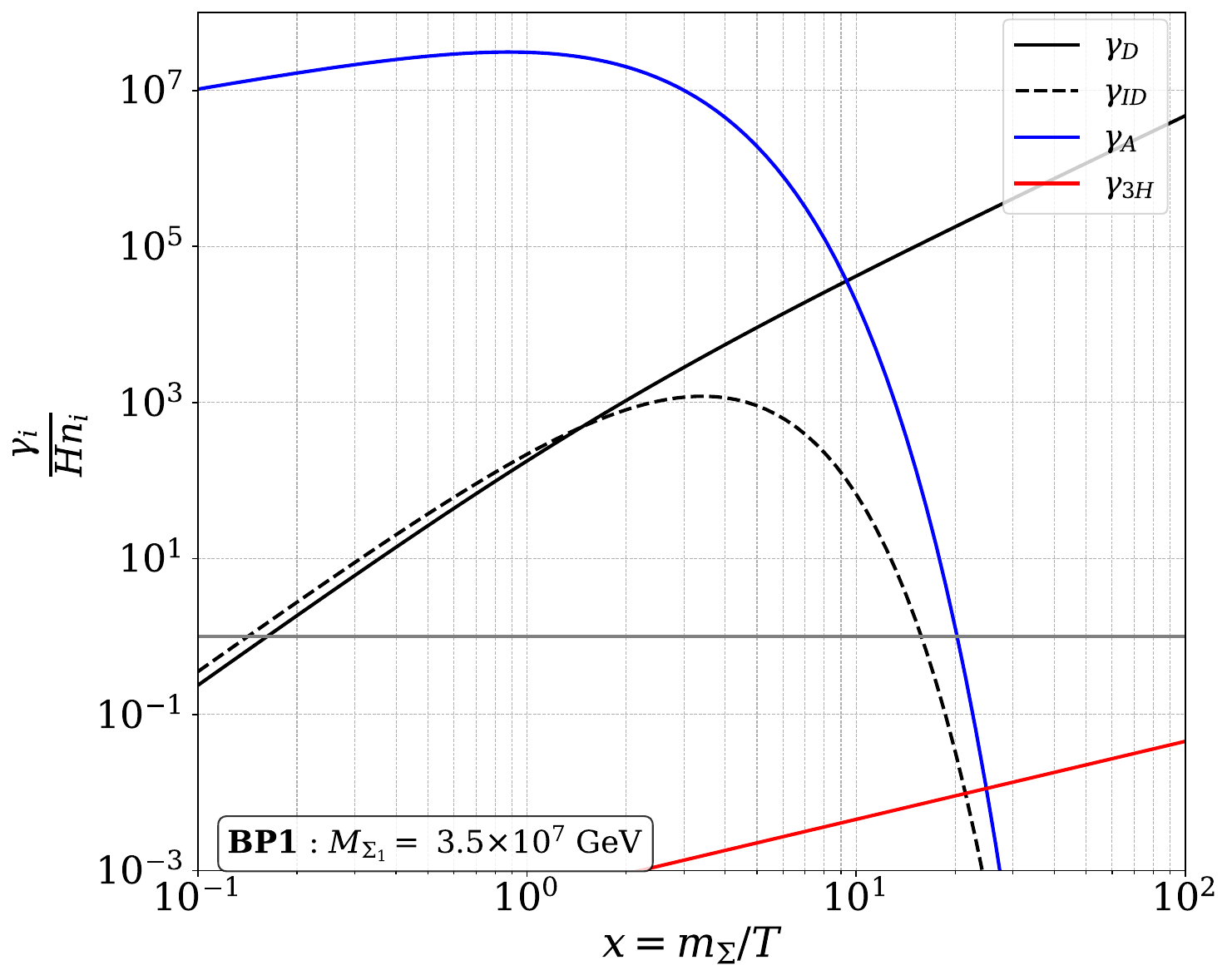}
\includegraphics[width=0.49\linewidth]{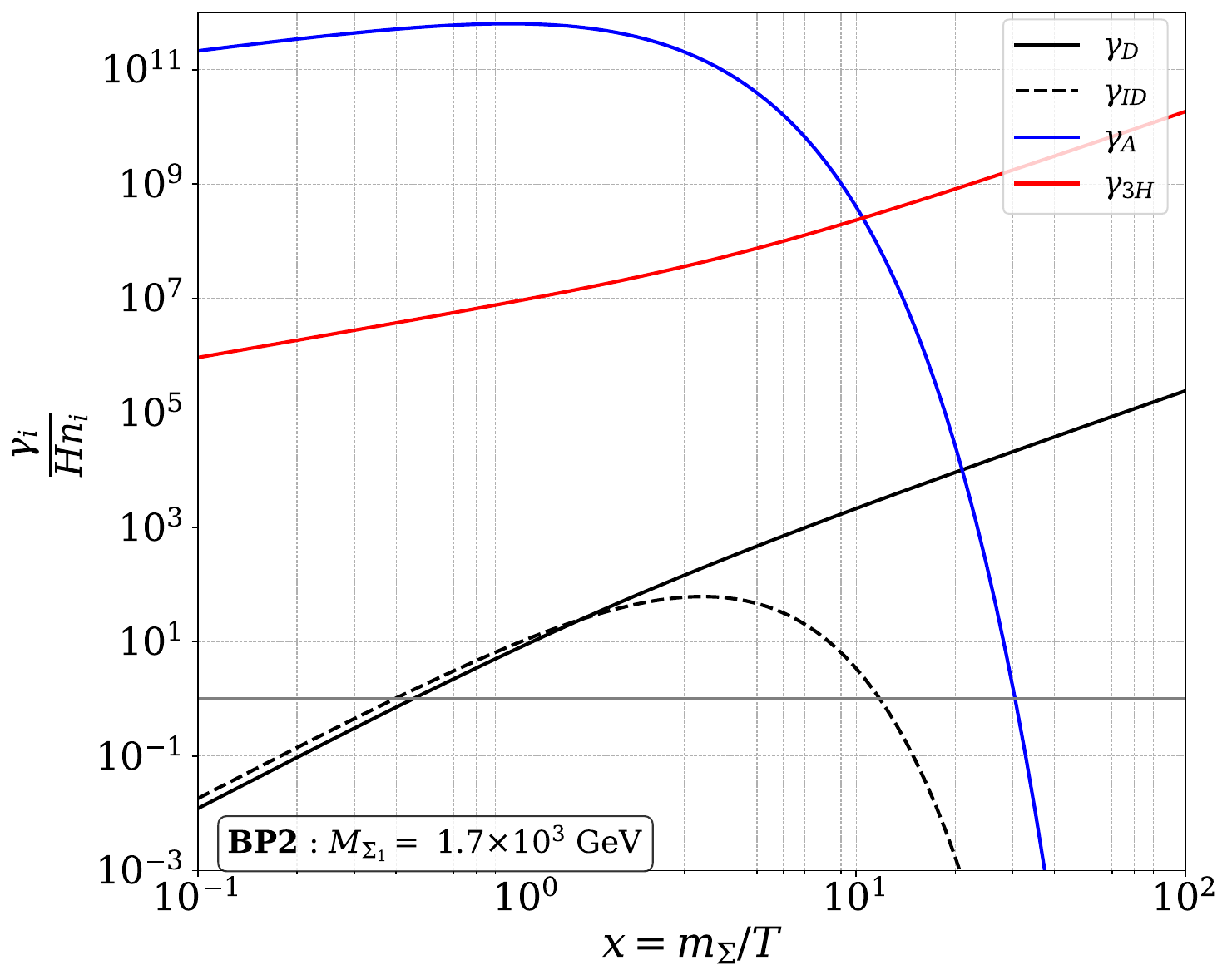}
\caption{Thermalization rates for gauge, Yukawa and scalar interactions relevant for the Boltzmann equations for two benchmark points indicated in Table~\ref{tab:bp}.}
\label{fig:rates}
\end{figure*}
The cosmological evolution of the asymmetries discussed above are determined by solving the following set of coupled Boltzmann equations (BEs)
   \begin{align}
    s H x \frac{dY_{\Sigma}}{dx} =& -2{\gamma}_{\Sigma} \, \left( \frac{Y_{\Sigma}}{Y_{\Sigma}^{\eq}}-1 \right) - 4 \, \gamma_{A} \left( \frac{Y_{\Sigma}^2}{{Y_{\Sigma}^{\eq}}^2} -1 \right)\,,  \label{eq:BE1}\\
    s H x \frac{dY_{\Delta \Sigma}}{dx}=& 2\gamma_\Sigma\, \left({\text{Br}_{\Phi}} \mathcal{X}_\Phi- {\text{Br}_{H}} \mathcal{X}_H \right)\,,\label{eq:BE2}\\
    s H x \frac{d Y_{\Delta \Phi}}{dx} =&  {\gamma_\Sigma} \, \left[ \vs \, \left( \frac{Y_{\Sigma}}{Y_{\Sigma}^{\eq}}-1 \right)  - 2 {\text{Br}_{\Phi}} \mathcal{X}_\Phi \right]\n\\
    &-\gamma_{3H}\left(\frac{Y_{\Delta \Phi}}{Y_\Phi^{\eq}}-\frac{3Y_{\Delta H}}{Y_H^\eq}\right)-2 \gamma_{\text{sub}} \left(\mathcal{X}_\Phi + \mathcal{X}_H \right)\,, \label{eq:BE3} \\
    s H x \frac{d Y_{\Delta H}}{dx} =& - \gamma_\Sigma \, \left[ \vs \, \left( \frac{Y_{\Sigma}}{Y_{\Sigma}^{\eq}}-1 \right)  - 2 {\text{Br}_{H}} \mathcal{X}_H \right]\n\\&+3\gamma_{3H}\left(\frac{Y_{\Delta \Phi}}{Y_\Phi^{\eq}}-\frac{3Y_{\Delta H}}{Y_H^\eq}\right)+2 \gamma_{\text{sub}} \left(\mathcal{X}_\Phi+\mathcal{X}_H\right) \,,\label{eq:BE4}\\
    s H x \frac{d Y_{\Delta L}}{dx} =& 2\gamma_\Sigma \, \left[ \vs \left(  \frac{Y_{\Sigma}}{Y_{\Sigma}^{\eq}}-1 \right) - {\text{Br}_{\Phi}} \mathcal{X}_\Phi - {\text{Br}_{H}} \mathcal{X}_H \right] \n\\
    &- 4 \gamma_{\text{sub}} \left(\mathcal{X}_\Phi+\mathcal{X}_H\right)\,, \label{eq:BE5}
\end{align} 
where $x \equiv M_{\Sigma_1}/T$ 
and
\begin{align}
    \mathcal{X}_\Phi &= \frac{ Y_{\Delta L}}{Y_L^{\eq} } +\frac{ Y_{\Delta \Phi}}{Y_\Phi^{\eq} } -\frac{Y_{\Delta \Sigma}}{Y_{\Sigma}^{\eq} }\,,\\
    \mathcal{X}_H &=\frac{Y_{\Delta \Sigma}}{Y_{\Sigma}^{\eq} }+\frac{ Y_{\Delta L}}{Y_L^{\eq} } -\frac{ Y_{\Delta H}}{Y_H^{\eq} }\,.
\end{align}
The equilibrium yield is given by
\begin{equation}
    Y_i^{\eq} = \frac{45\,g_i}{4\pi^4 \,g_\ast} x_i^2\, K_2(x_i)\,,
\end{equation}
where $x_i\equiv M_i/T$ and $K_n(x)$ is the modified Bessel function of the second kind of order $n$.

The first BE~\eqref{eq:BE1} takes into account the evolution of $\Sigma \equiv \Sigma_1 + \bar{\Sigma}_1$ which includes its decay, inverse decay and scattering processes. The rest of the BEs~\eqref{eq:BE2}-\eqref{eq:BE5} track the evolution of asymmetries in different species. The decay and scattering rate densities are given by
\begin{align}
    \gamma_{a \r ij} &= n_a^\eq \frac{K_1(m_a/T)}{K_2(m_a/T)}\Gamma_{a \r ij}\,,\n\\
    \gamma_{ab \r ij}&=\frac{T}{64\pi^4}\int_{s_{\text{min}}}^\infty ds\, \sqrt{s}\, \hat{\sigma}(s)\,K_1\left(\frac{\sqrt{s}}{T}\right)\,.
\end{align}
Here $\hat\sigma$ is the reduced cross-section and $s_{\rm min}=(m_a+m_b)^2$.  
Analogous to the case of Type-II seesaw~\cite{Hambye:2005tk}, not all of these equations are independent as there exists a sum rule from hypercharge conservation, which leads to the following relation among the asymmetries in various species
\begin{equation}
    2\Delta\Sigma + 3\Delta \Phi + \Delta H - \Delta L = 0\,.\label{eq:asym-conservation}
\end{equation}
Below, we discuss the effects of various terms included in the BEs above.\\

\noindent\textbf{Gauge interactions:} Due to their non-zero hypercharge, the fermion triplets in this model have stronger gauge interactions than that of Type-III seesaw. The gauge interactions include $\Sigma \bar{\Sigma} \leftrightarrow f\bar{f}, HH^{\ast}, GG'$ where $G\,(f)$ denotes gauge bosons (fermions). The total reduced cross-section for these annihilations that tend to suppress the $\Sigma$ population is given by~\cite{Cirelli:2007xd}
\begin{align}
    \hat{\sigma}_A =& \frac{3(2g^2+g'^2)^2}{2\pi}
    \left[2\beta(\beta^2-2)-(\beta^4-3)\ln\left(\frac{1+\beta}{1-\beta}\right)\right] ,
\end{align}
with $\beta \equiv \sqrt{1-4/y}$ and $y \equiv s/M_\Sigma^2$. The reaction density can then be written in terms of $\hat{\sigma}_A$ as
\begin{align}
    \gamma_A = \frac{M_\Sigma^4}{64\pi^4}\int_{y_{\text{min}}}^\infty dy \sqrt{y}\, \frac{K_1(x\sqrt{y}) \, \hat{\sigma}_A}{x}\,.
\end{align}
As mentioned above, these gauge scatterings/annihilations are initially not in thermal equilibrium for $T \gg M_\Sigma$, and are doubly Boltzmann suppressed at late times, $T \ll M_\Sigma$. The strength of these interactions is maximal at $T \sim M_\Sigma$. This behavior can be seen from the blue curves in Fig.~\ref{fig:rates}. For the triplet with hypercharge 1, it can be checked that for $M_\Sigma < 10^{15}$ GeV, the triplets always thermalize irrespective of their initial abundance, hence, they play an important role in the evolution of $\Sigma$ abundance. However, they do not appear in any of the asymmetry equations as they always involve pair annihilations.\\
\begin{figure}[!t]
    \centering
    \includegraphics[width=0.95\linewidth]{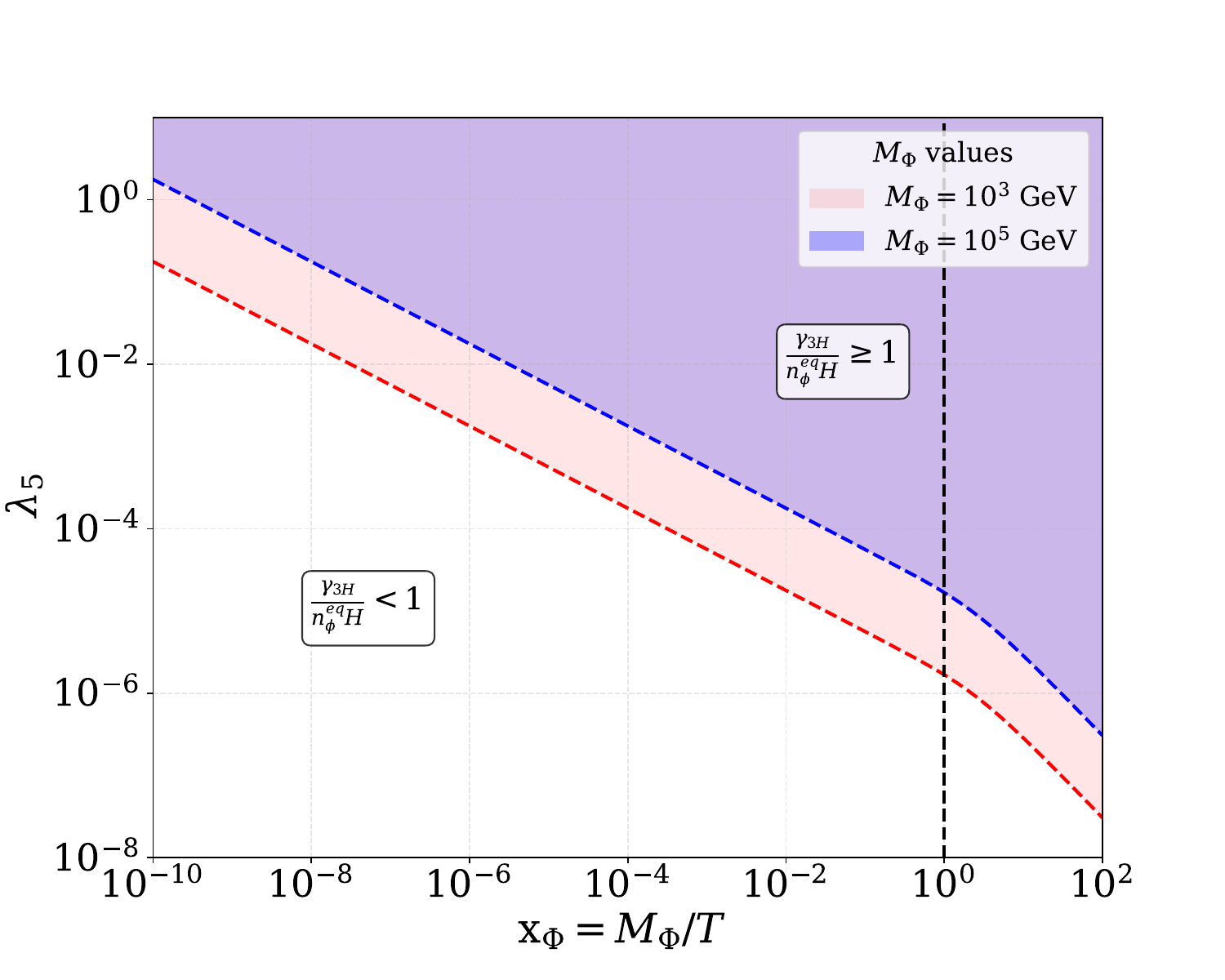}
    \caption{The region in the $(x_\Phi,\lambda_5)$ plane where the scalar quartic interactions are in equilibrium (shaded regions) for different values of $M_\Phi$. The vertical line is for $T=M_\Phi$.}
    \label{fig:phieq}
\end{figure}

\noindent\textbf{Yukawa interactions:} The effect of Yukawa interactions that leads to the decays of $\Sigma$ is incorporated in $\vs$, and $\gamma_\Sigma$ which includes the total decay width of $\Sigma$. These interactions compete against the gauge interactions; if $\gamma_A/(n_\Sigma H)>1$ and $\gamma_\Sigma < \gamma_A$, the gauge interactions dominate and the triplets annihilate before decaying, thus significantly suppressing the efficiency of leptogenesis. At $T< M_\Sigma$, the decays have a single Boltzmann suppression in contrast with the double suppression of the gauge interactions. Hence, if $\gamma_\Sigma \geq 4\gamma_A$, the decays dominate and the triplets will decay before annihilating, making leptogenesis more efficient. We find that $\gamma_\Sigma \geq 4\gamma_A$ around $x \equiv x_A \sim 10$ for $\mathcal{O}(0.1)$ Yukawa couplings. 

Since the Yukawa couplings in the model are not suppressed by neutrino mass, and can take larger values than in canonical seesaws, it is easy to satisfy the condition $\gamma_\Sigma \geq 4\gamma_A$ for $\abs{Y_H}, \abs{Y_\Phi} \gtrsim 0.1$. In this case, the $\gamma_A$ term can be neglected in the BEs, and the suppressed efficiency of leptogenesis is mainly due to the washout from inverse decays that scale as $\abs{y}^2$, which become suppressed at $T < M_\Sigma$. The inverse decays are taken into account by the terms in the square brackets not multiplied by $\vs$ in Eqs.~\eqref{eq:BE3}-\eqref{eq:BE5}. The efficiency is enhanced if these inverse decays also decouple around $x_A$. The decay and inverse decay rate are shown by black solid and dashed curves, respectively, in Fig.~\ref{fig:rates} for our choice of benchmark points. 

Next, we discuss the $\Delta L =2$ scatterings involving the Yukawa interactions. The vector-like nature of $\Sigma$ allows only $\Delta L=2$ scattering $L\Phi \leftrightarrow \bar{L}H$ and its conjugate process. The on-shell rate of these processes has already been accounted for via the inverse decays and decays in the BEs; therefore, only the on-shell or real intermediate state (RIS)-subtracted part of these rates needs to be tracked. We define $\gamma_{\text{sub}}= \gamma_s + \gamma_t$ corresponding to the $s$- and $t$-channel processes. The complete expressions for these can be found in Appendix~\ref{app:rates}. These rates are not only suppressed by $\abs{y}^4$, but also fall off at $x \gg 1$, therefore their effects are sub-dominant compared to the decays/inverse decays for $\abs{y} \lesssim \mathcal{O}(0.1)$, and can be neglected for $M_\Sigma \gtrsim \mathcal{O}(\text{TeV})$, as discussed above.  \\

\noindent\textbf{Scalar interactions:} The $\lambda_5$ term in Eq.~\eqref{eq:potential} required for neutrino masses leads to the three-body decays $\Phi \r HHH$ as well as $2 \leftrightarrow 2$ scatterings such as $\Phi H^\ast \leftrightarrow HH$. The decay rate for $\Phi \r HHH$ in the limit $M_\Phi \gg M_H$ can be written as
\begin{align}
    \Gamma_{\Phi\to 3H} \simeq & \frac{3\lambda_5^2}{256\,\pi^3}M_\Phi\,.
\end{align}
The reduced scattering cross-section for the contact interaction in the limit $s \gg M_\Phi^2,M_H^2$ is given by
\begin{align}
    \hat{\sigma}_{\Phi H^\ast \r HH} \simeq \frac{9\lambda_5^2}{16\pi}\,.
\end{align}
Hence, we define
\begin{align}
    \gamma_{3H} &\equiv \gamma_{\Phi\to 3H} + \gamma_{\Phi H^\ast \r HH}\nonumber\\
    &=\frac{3\lambda_5^2\, M_\Phi^4}{128\pi^5\, x_\Phi^2}\, \left[x_\Phi K_1\left(x_\Phi\right)+3 K_2(x_\Phi)\right]\, .
\end{align}
The $\gamma_{3H}$ term appears only in the BEs for $\Delta \Phi$ and $\Delta H$, where the factor of $3$ in the BE for $\Delta H$ appears as these processes change $H$ number by 3 units. In Fig.~\ref{fig:phieq}, we show the region where these interactions are in equilibrium in the $\lambda_5$ and $x_\Phi$ plane for different values of $M_\Phi$. The thermalization rate for these interactions is shown by red solid lines in Fig.~\ref{fig:rates}.
\begin{figure*}[!t]
    \centering
    \includegraphics[width=0.49\linewidth]{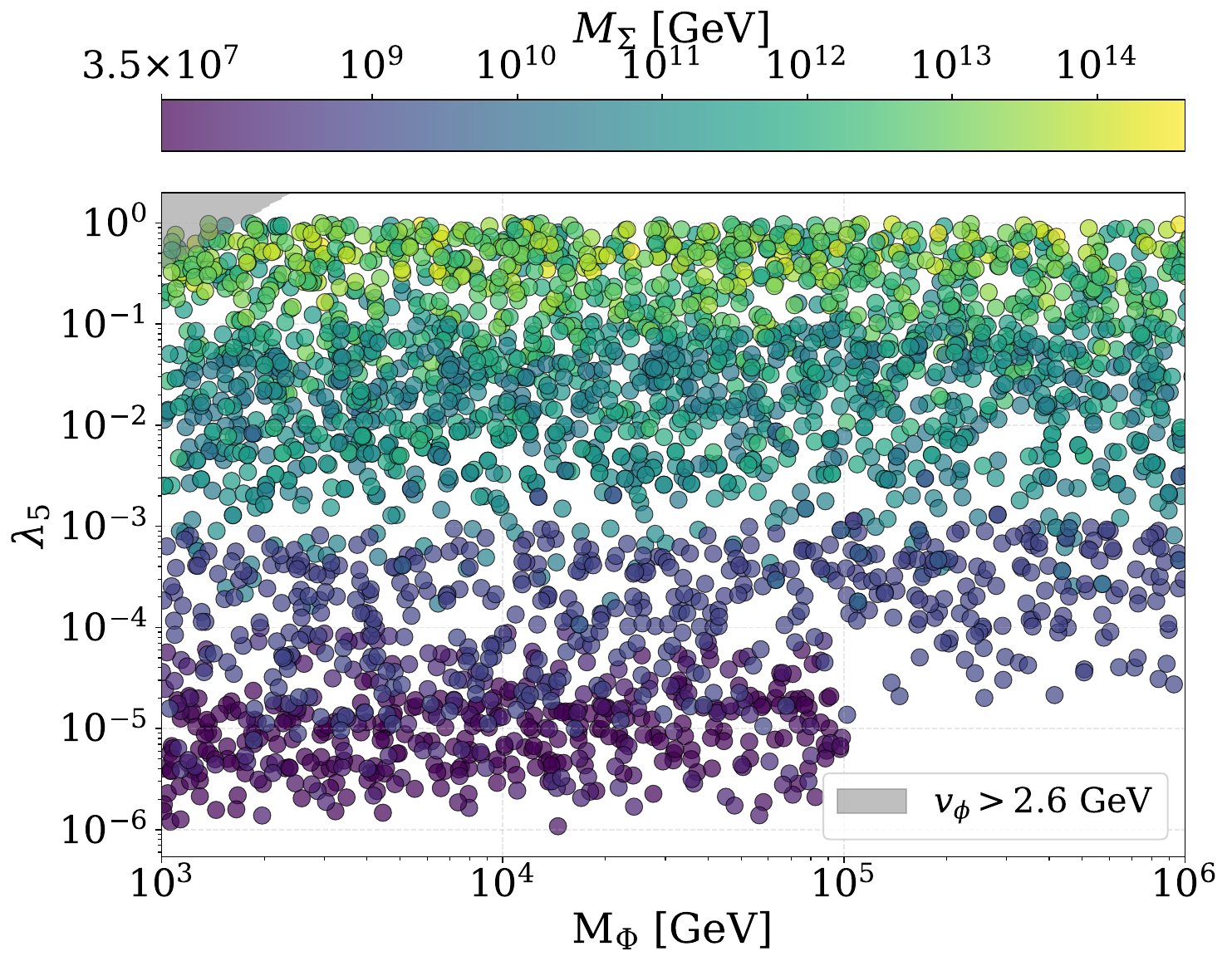}
    \includegraphics[width=0.45\linewidth]{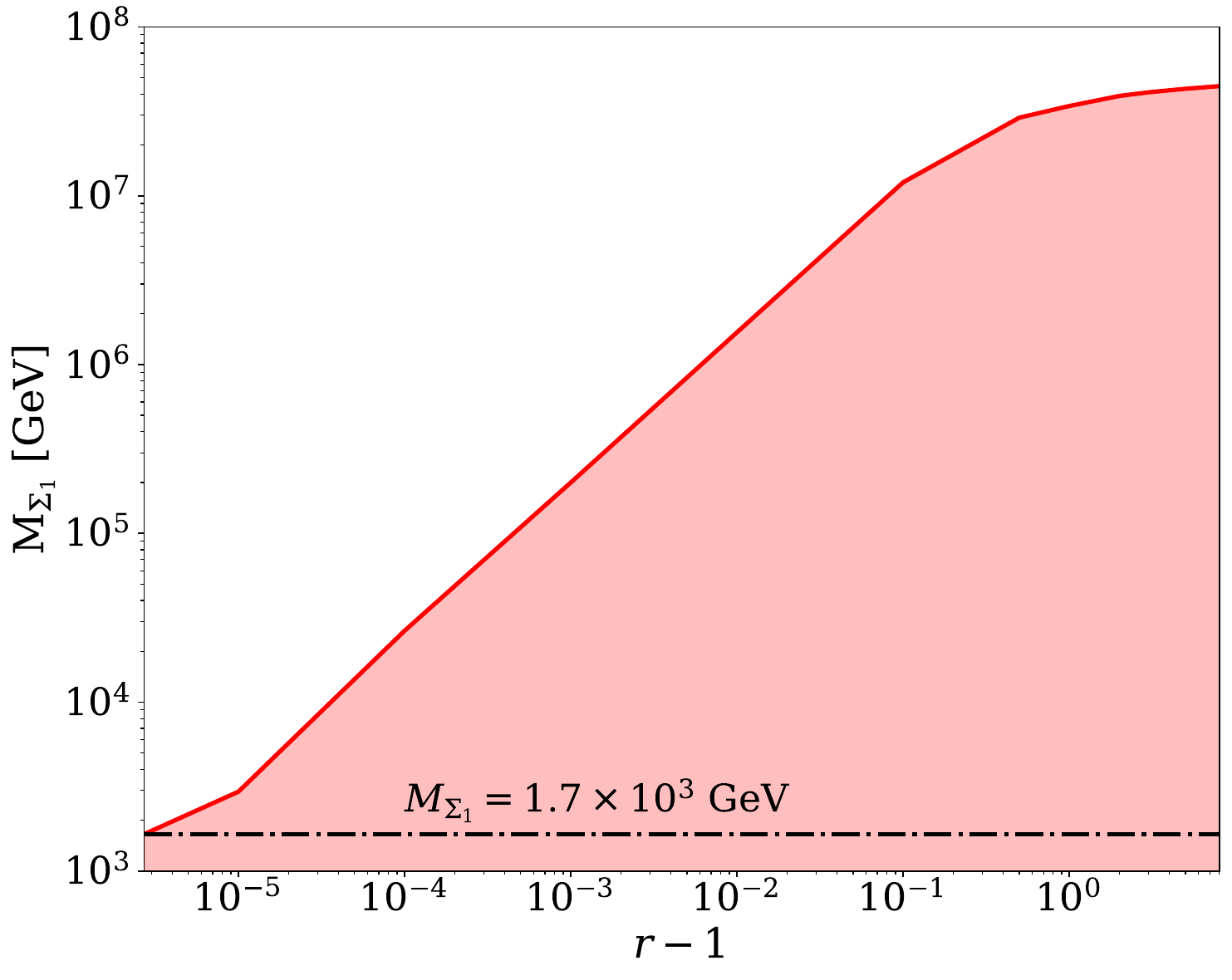}
    \caption{\textit{Left:} The viable parameter space for hierarchical leptogenesis in the BNT model in the $(M_\Phi,\lambda_5)$ plane, where each point corresponds to a different value of $M_\Sigma$ as shown in the color bar. The gray region in the top-right corner is excluded from the constraints on the $\rho$ parameter. \textit{Right:} The dependence of the lower bound on $M_\Sigma$ on the hierarchy among the triplet masses. Leptogenesis is not viable in the shaded region.}
    \label{fig:scan}
\end{figure*}
\section{Results}\label{sec:results}

The asymptotic lepton asymmetry generated in $\Sigma$ decays can be determined by numerically solving the coupled BEs \eqref{eq:BE1}-\eqref{eq:BE5}. Let us first discuss the implications for different asymmetry equations. If we ignore all the contributions from scalar interactions and the washout from inverse decays, the set can be simplified by ignoring the scalar asymmetries to only solve for $Y_\Sigma, Y_{\Delta \Sigma}$ and $Y_{\Delta L}$. In this case, it is easy to see that $Y_{\Delta L} = Y_{\Delta \Phi} - Y_{\Delta H}$, and $Y_{\Delta H} = -Y_{\Delta \Phi}$, as all these asymmetries have the same source. 

Next, let us include the $\gamma_{3H}$ term which ties the scalar asymmetries together. If these interactions are in equilibrium at $T \sim M_\Sigma$, then the asymmetry in $\Phi$ will be quickly washed out. Conversely, if they remain out of equilibrium at all times, they can be neglected. In Fig.~\ref{fig:phieq}, we see that these interactions are always in equilibrium at $T \sim M_\Phi$ for $\lambda_5> 10^{-6}$; therefore, in order to track the lepton asymmetry correctly, we must carefully account for these interactions and track the scalar asymmetries.\footnote{Since the $\gamma_{3H}$ interactions come into equilibrium already when the sphalerons are active, we include their effect to determine the modified sphaleron conversion factor in the model; see Appendix~\ref{app:Sphaleron}.} Similarly, including the inverse decay contribution encoded in $\mathcal{X}_{\Phi, H}$ also relates the scalar asymmetries to the lepton asymmetry.

We find that including the $\gamma_{3H}$ processes can change $Y_{\Delta L}$ by $\mathcal{O}(1)$ factors for the range of parameters we work with. This is due to the fact these interactions suppress the $\Phi$ population and thus modify the washout from inverse decays involving $\Phi$. In other words, $\Phi$ produced in the decays of $\Sigma$ can quickly decay to the SM Higgs or scatter before it can participate in the inverse decay contributing to washout, thus slightly enhancing the final lepton asymmetry.

In order to analyze the parameter space where leptogenesis is viable, we numerically solve the BEs for a random scan of the following set of free parameters:
\begin{equation*}
    \{M_\Sigma, M_\Phi, \lambda_5,\theta_{ij}, r\equiv M_{\Sigma_2}/M_{\Sigma_1}\}\,,
\end{equation*}
as the Yukawa couplings are already fixed by the GCI parameterization.  The ranges  over which the  parameters of interest are scanned are given in Table~\ref{tab:scan_ranges}. We select those points that lead to $Y_{\Delta L}^{0} = (1.92 \pm 0.04) \times 10^{-10}$. The parameter space where leptogenesis can work in the model is shown in Fig.~\ref{fig:scan}. 
\begin{table}[!htb]
    \begin{tabular}{l| l}
        \toprule
           Parameters & Range\\
           \hline
           $M_{\Sigma_1}$  & $[10^3 \,, \, 10^{15}]$ GeV\\
           $r$ & $[2\,, 10]$\\
           $M_\Phi$& $[10^3\,,\, 10^6]$ GeV\\
           $\lambda_5$ & $[10^{-7} \,, \, 1]$\\
           $\theta_{ij}$ & $[10^{-5}\, ,\, 10^5]\times \exp\left(i~[0, 2\pi]\right)$\\
           \bottomrule
    \end{tabular}
    \caption{Summary of the model parameters and their corresponding variation ranges explored in the numerical scans.}
    \label{tab:scan_ranges}
\end{table}
The scale of $M_\Sigma$ above which leptogenesis becomes viable depends on the hierarchy among the triplet masses, we show this dependence in the right plot of Fig.~\ref{fig:scan}. Note that for $r \simeq 1$, we enter the quasi-degenerate regime which requires a different approach, as we discuss in the next section. 

For $r=2$, we find that leptogenesis is viable for
\begin{equation}
    M_{\Sigma_1} \gtrsim 3.5 \times 10^{7}~\text{GeV} , 
\end{equation}
while for larger values of $r$, the lower bound goes up slightly, as shown in Fig.~\ref{fig:scan} right panel. 
A representative set of  values of other parameters for this fixed value of $M_\Sigma$ are shown in Table~\ref{tab:bp}, and we label the set as BP1.
\begin{table*}
    \begin{tabular}{c c c c c c c c c c}
        \toprule
          & $M_{\Sigma_1}$ [GeV] & $r$ & $M_\Phi$ [GeV]& $\lambda_5$ & $v_\Phi$ [GeV] & $\theta_{12}$ & $\theta_{13}$ &  $\theta_{14}$ & $\theta_{34}$ \\
          \midrule
          BP1 & $3.5 \times 10^7$ & $2$ & $1000$ & $10^{-5}$ & $7.4 \times 10^{-5}$ & $-0.003-0.005~i$ & $0.001 $ & $0.241-0.179~i$ & $-1.56 + 0.018~i$\\
          BP2 & $1.7 \times 10^3$ & $\simeq 1$ & $1000$ & $0.01$ & $0.074$ & $-0.003-0.005~i$ & $0.001 $ & $0.241-0.179~i$ & $-1.56 + 0.018~i$\\
         \bottomrule
    \end{tabular}
    \caption{Set of benchmark points associated with the lowest possible scale for $M_{\Sigma_{1}}$ above which one can successfully reproduce the baryon asymmetry via leptogenesis in the BNT model. BP1 and BP2 correspond to the hierarchical and resonant case respectively. The remaining angles $\theta_{23}$ and $\theta_{24}$ in the GCI parameterization are set to zero here.}
    \label{tab:bp}
\end{table*}
The corresponding Yukawa matrix structure comes out to be
\begin{align}
Y_{\Phi} & = \begin{pmatrix}
    -0.008 -0.015~i & -0.98-2.2~i \\
   0.026+0.080~i  & 5.38-1.86~i\\ 
    0.041 + 0.067~i & 4.76 + 1.77~i   
\end{pmatrix} \times 10^{-2} \, , \nn \\
Y_{H} & = \begin{pmatrix}
    0.002 + 0.020~i & -0.98-1.2~i\\
    0.052 -0.062~i & 5.55 -1.86~i\\
    0.036 - 0.063~i & 4.91 +1.77~i  
\end{pmatrix} \times 10^{-2} \, .
\end{align}
First of all, we see that $Y_\Phi \sim Y_H$, consistent with the discussion above for maximizing the $CP$ asymmetry. Second, the structure exhibits a strong hierarchy among the Yukawa couplings of $\Sigma_1$ and $\Sigma_2$, $\abs{Y_{i1}} \ll \abs{Y_{i2}}$, implying that the inverse decays involving $\Sigma_1$ are further suppressed; this pattern is also seen in other low-scale leptogenesis scenarios~\cite{Racker:2024fpn}. Moreover, the similar values of the Yukawa couplings along with a small $\lambda_5$ for this BP suggest that giving $\Phi$ a lepton number, ${\sf L} = -2$ is more natural in the hierarchical case.

\begin{figure*}[!t]
    \centering
        \includegraphics[width=0.49\linewidth]{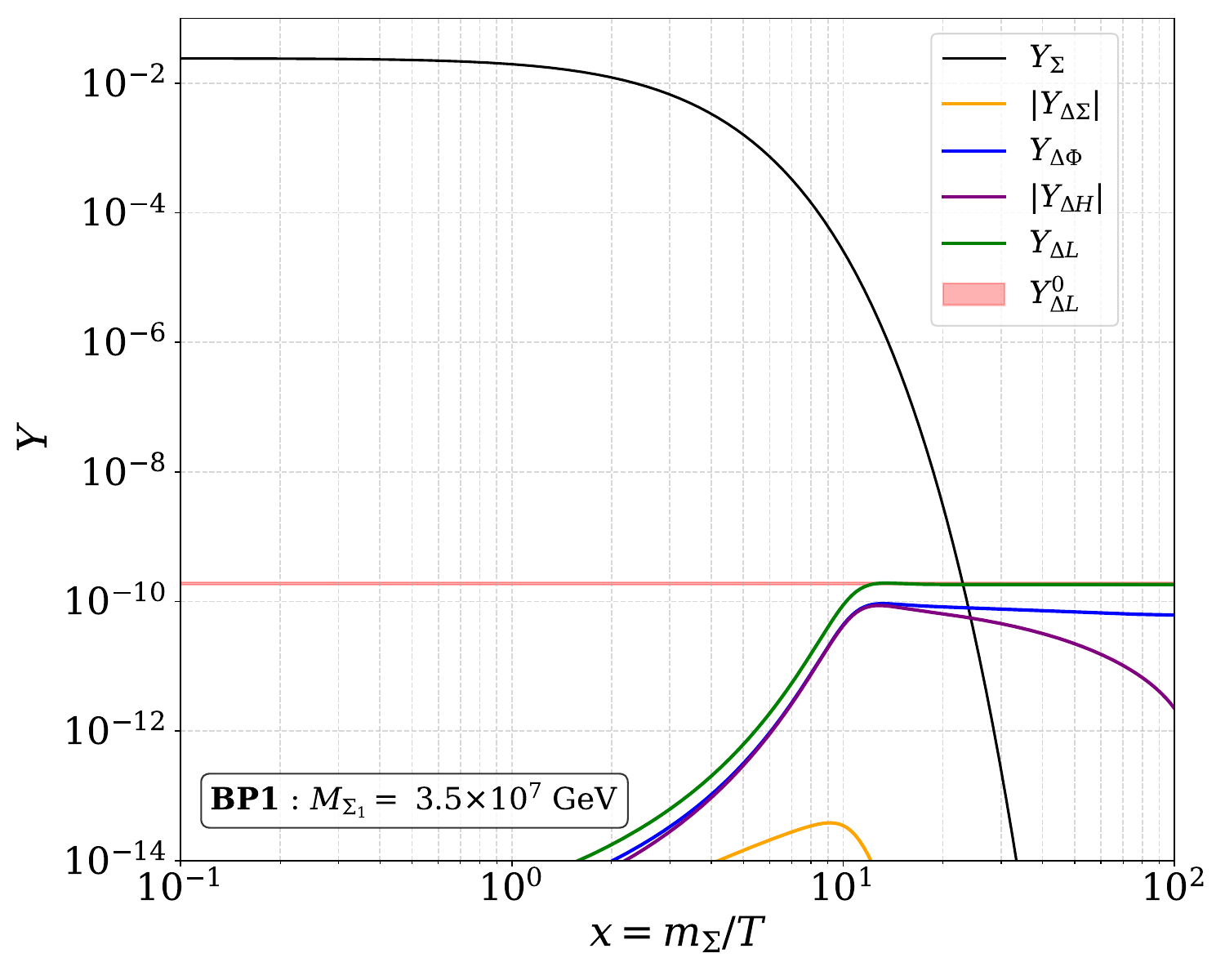}
        \includegraphics[width=0.49\linewidth]{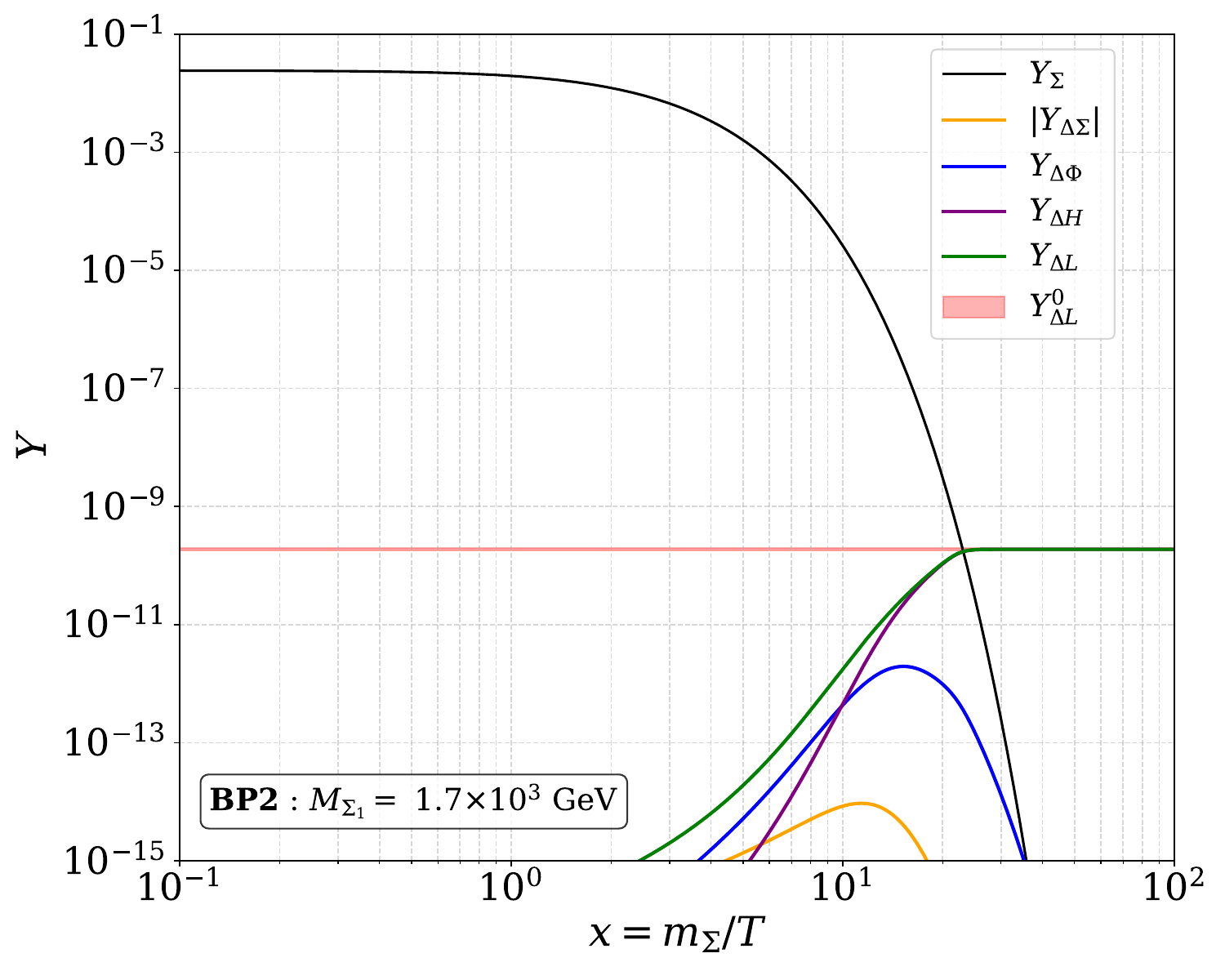}
    \caption{Evolution of abundances and asymmetries in the hierarchical (left) and quasi-degenerate regime (right) for our two BP choices shown in Table~\ref{tab:bp}. The solid red line corresponds to the required asymmetry today $Y^0_{\Delta L} = (1.92 \pm 0.04)\times 10^{-10}$. }
    \label{fig:abundances}
\end{figure*}

We also  show the evolution of abundances and asymmetries for BP1 in the left panel of Fig.~\ref{fig:abundances} in order to understand the dynamics. It can be seen that the lepton and the scalar asymmetries (green, blue and violet curves) get dominantly produced at $x > 1$, once the decays of $\Sigma$ go out of equilibrium and its abundance gets suppressed (black curve). The solid orange line shows the evolution of the asymmetry in $\Sigma$, with the lepton asymmetry freezing around the peak of the orange curve, after which the $\Sigma$ asymmetry decays away. On the other hand, the scalar interactions will deplete the asymmetry in $\Phi$ once they come into equilibrium, as seen from the downward gradient of the blue curve. Notice that $Y_{\Delta L} $ and $Y_{\Delta \Phi}$ curves are following each other closely, therefore, in order to satisfy Eq.~\eqref{eq:asym-conservation}, $Y_{\Delta H} = Y_H - Y_{H^{\ast}}$ comes out to be negative. Indeed, our numerical solution agrees with this, and we thus plot $\left| Y_{\Delta H }\right|$ in violet. Hence, for $x > 1$, the downward slope of the violet curve corresponds to the increase in the $H$ population coming from $\Phi \to 3H$ decays.

\section{Low-Scale Leptogenesis }\label{sec:resonant}
The results from the hierarchical scenario show that leptogenesis is viable a few orders above the TeV scale, thus undermining the original motivation of the BNT model to realize a realistic neutrino mass generation mechanism along with interesting TeV-scale phenomenology. To alleviate this tension, we now consider the possibility of lowering the leptogenesis scale via resonant enhancement of the $CP$ asymmetry by considering a quasi-degenerate spectrum of $\Sigma$, similar to the resonant leptogenesis in canonical seesaw models~\cite{Pilaftsis:2003gt, Dev:2017wwc}. For the resonant enhancement, we require 
\begin{eqnarray}
    \Delta M_\Sigma \simeq \frac{\Gamma_{\Sigma_2}}{2} \ll M_{\Sigma_{1,2}}\, .
\end{eqnarray}
It can be seen that in the limit $M_{\Sigma_{1}} \simeq M_{\Sigma_{2}}$, the self energy loop function in Eq.~\eqref{eq:loopfn} diverges. Hence, the loop function must be regulated by including the $\Sigma_2$ decay width, so that we have~\cite{BhupalDev:2014pfm}
\beqa
f_s^{\text{reg}}(z_\Sigma,z_\Gamma) & = & \frac{(1-z_\Sigma)z_\Gamma}{(1-z_\Sigma)^2 + z_\Gamma^2}\, ,\label{eq:self-loog-regulated}
\eeqa
where $z_\Sigma = M^2_{\Sigma_2}/M_{\Sigma_1}^2$ and $z_\Gamma = \Gamma_{\Sigma_2}/M_{\Sigma_1}$. The $CP$ asymmetry in the resonant scenario can be written as 
\begin{align}\label{eq:cpres}
    \varepsilon_H &= \frac{1}{2048\pi^3}\frac{\text{Im}[\mathbf{Y}]{f}_{s}^{\text{reg}} \left(z_\Sigma,z_{\Gamma}\right) \, M_{\Sigma_1} M_{\Sigma_2}}{\Gamma_{\Sigma_1} \, \Gamma_{\Sigma_2}}\,,\n\\
    \varepsilon_\Phi &= \frac{1}{2048\pi^3}\frac{\text{Im}[\mathbf{Y}^\dagger]{f}_{s}^{\text{reg}} \left(z_\Sigma,z_{\Gamma}\right) \, M_{\Sigma_1} M_{\Sigma_2}}{\Gamma_{\Sigma_1} \, \Gamma_{\Sigma_2}}\,,
    \end{align}
and can be as large as $\mathcal{O}(1)$. This enhancement thus allows us to have leptogenesis at the TeV scale. We find that for $r\simeq 1$, $M_{\Sigma_1} \geq 1.7$ TeV (see Fig.~\ref{fig:scan} right panel), similar to that for Type-II and Type-III seesaw leptogenesis~\cite{Strumia:2008cf}, thus restoring the original motivation for the BNT model as providing a new neutrino mass generation mechanism at the TeV scale, while simultaneously explaining the BAU. 

The representative values of other parameters corresponding to this value of $M_\Sigma$ are labeled as BP2 in Table~\ref{tab:bp}. The corresponding Yukawa matrices are given by
\begin{eqnarray}
Y_{\Phi} & = \begin{pmatrix}
    -0.013 - 0.024~i &   -1.13-2.52~i\\ 
        0.041 + 0.13~i &  6.22-2.16~i\\
        0.064 +0.11~i & 5.51 + 2.04~i 
\end{pmatrix} \times 10^{-5},\nn \\
Y_{H} & = \begin{pmatrix}
    0.003 + 0.032~i &  -1.13 -1.38~i\\
    0.082 -0.098~i & 6.42 -2.15~i\\
    0.058 -0.099~i &  5.68 +2.04~i
\end{pmatrix} \times 10^{-5}\,.
\end{eqnarray}
Once again, the Yukawa matrices are of similar order$Y_H \sim Y_\Phi$ and $\abs{Y_{i1}}\ll \abs{Y_{i2}}$ like the hierarchical case above. However, notice that they are much smaller $\sim \mathcal{O}(10^{-5})$ compared to BP1. This decrease in Yukawas is instead compensated by a larger value of $\lambda_5 \simeq \mathcal{O}(10^{-2})$. Therefore, in this scenario, it is natural to shift LNV directly into the Yukawa sector instead of the scalar sector. 

The evolution of the abundances and asymmetries corresponding to BP2 are shown in the right panel of Fig.~\ref{fig:abundances}, and can be contrasted with the hierarchical case shown in the left panel. The consequence of large $\lambda_5$ is immediately seen by the rise and decline of the asymmetry in $\Phi$, while the lepton asymmetry freezes in. For BP2, as $\gamma_{3H}$ is always in  equilibrium during the evolution, $Y_{\Delta H} $ is positive in contrast to BP1. In addition, for $x>1$, as $Y_{\Delta \Phi}$ quickly goes to zero,  the populations of $Y_{\Delta H}$ and $ Y_{\Delta L}$ become exactly the same which once again validates Eq.~\eqref{eq:asym-conservation}.  

Moreover, models with low-scale seesaws are in general subjected to constraints from lepton flavor violation, and in this case also Higgs lepton flavor violating couplings, see Ref.~\cite{Herrero-Garcia:2016uab}. This is attributed to the presence of a dimension-6 operator that is obtained after integrating the heavy fermions out, yielding a non-unitary PMNS matrix. Such constraints typically place limits on the Yukawa couplings of the form~\cite{Giarnetti:2023osf}
\begin{align}
    Y_{\beta i}^\ast Y_{\alpha i}\left(\frac{{\rm TeV}}{M_\Sigma}\right)^2 \leq \begin{cases}
        0.0004~\text{for}~\alpha\beta = \mu e\\
        0.24~\text{for}~\alpha\beta = \tau e\\
        0.28~\text{for}~\alpha\beta = \tau\mu\,,
    \end{cases}
\end{align}
which is easily satisfied by our choice of BP2. Further, the collider phenomenology for the TeV-scale BNT model has already been discussed in Ref.~\cite{Bambhaniya:2013yca, 
Ghosh:2017jbw,Ghosh:2018drw, Pan:2019wwv,Giarnetti:2023dcr, Chakraborty:2025kcl} which applies to our resonant case as well. 
We show the values of $v_\Phi$ for our BPs in Table~\ref{tab:bp}. In the hierarchical case, as $\lambda_5$ is suppressed, we get $v_\Phi = 74$ keV, implying that the dominant decay channel is to dileptons, whereas for the resonant case, we get $v_\Phi=74$ MeV and the dominant decays will be to a pair of gauge bosons.

\section{Conclusions}\label{sec:conc}

In this work, we have studied leptogenesis in the Babu–Nandi–Tavartkiladze (BNT)  model with fermion triplets and a scalar quadruplet. In this model, neutrino masses are generated at the tree level via a dimension-7 effective operator, as well as at the one-loop level via the dimension-5 Weinberg operator. Although this model was proposed many years ago and its collider phenomenology has been extensively studied, to the best of our knowledge, this is the first detailed study of leptogenesis in the BNT model and, more generally, in a framework where both fermionic and scalar multiplets larger than $SU(2)_L$ triplets participate in neutrino mass generation and leptogenesis.

The presence of vector-like fermions and two independent Yukawa couplings makes this scenario distinct from canonical seesaw-based leptogenesis. Whereas in canonical Type-I leptogenesis the heavy Majorana mass provides the source of lepton-number violation, the original BNT realization contains vector-like fermions with Dirac masses, while lepton number is broken by the scalar interaction. Moreover. The small induced vacuum expectation value of the scalar quadruplet allows significantly larger Yukawa couplings while reproducing the observed neutrino masses. This enhances the CP asymmetry and lowers the scale required for successful leptogenesis compared with canonical seesaw scenarios.

We also consider the alternative realization discussed in  Ref.~\cite{Bambhaniya:2013yca}, in which lepton-number violation originates in the Yukawa sector. We show that both realizations can successfully account for the observed baryon asymmetry, with the Yukawa-induced realization being particularly well suited for resonant leptogenesis.

We find that for a hierarchical spectrum of triplet masses $M_{\Sigma_2} \geq 2M_{\Sigma_{1}}$, leptogenesis is viable for $M_{\Sigma_1}\gtrsim {\cal O}(10^7)$ GeV which, although lower than the DI bound of $10^9$ GeV in canonical seesaw, is still beyond the experimental reach, thus undermining the original TeV-scale testability of the model. However, a quasi-degenerate spectrum of fermion triplets allows us to resonantly enhance the asymmetry production and lowers the viable scale down to $\mathcal{O}$(TeV). While we do not consider flavor effects~\cite{Dev:2017trv} in our analysis, which can indeed be important for the scales mentioned above, we expect the bound in the hierarchical case to be further lowered by about an order of magnitude, analogous to the case of Type-III leptogenesis with an additional Higgs doublet getting a small VEV~\cite{Vatsyayan:2022rth}. On the other hand, the bound in the resonant case is extremely close to the absolute lower bound on leptogenesis with triplet fermions and scalars evaluated after considering Sommerfeld corrections~\cite{Strumia:2008cf}, and is therefore not expected to be lowered further once the flavor effects are included. 

The results of this work can be analogously applied to other neutrino mass models with higher-dimensional operators  involving multiplets beyond the triplets. The viable leptogenesis scale could be similar or even smaller by a few orders of magnitude as they would be free from the constraints on the SM Higgs, unlike the BNT model. Future searches for multi-charge fermions and scalars at the colliders as well as probes of lepton flavor violation at low scale can shed some light on the parameter space of these higher-dimensional operator-based neutrino mass models.

\begin{acknowledgments}
    We thank Kaladi Babu, Sudip Jana and Shaikh Saad for useful discussions. The work of BD was supported in part by the U.S. Department of Energy under grant No. DE-SC0017987 and by a Humboldt Fellowship from the Alexander von Humboldt Foundation. SG acknowledges the J.C.~Bose Fellowship (JCB/2020/000011) of the Anusandhan National Research Foundation, and the Department of Space, Government of India. The work of DP is funded by the Indian Space Research Organisation (ISRO) under grant STC/PHY/2024427Q. DV is supported by Canada First Research Excellence Fund through the Arthur B. McDonald Canadian Astroparticle Physics Research Institute, and a Subatomic Physics Discovery Grant (individual) from the Natural Sciences and Engineering Research Council of Canada. BD thanks the organizers of DTP-TALENT 2026 at ECT$^*$, Trento for  local
    hospitality during the final stages of this work.
\end{acknowledgments}

\appendix
\section{Decay and Scattering Rates}\label{app:rates}

The total decay width of $\Sigma$ can be written as
\begin{align}
\Gamma_{\Sigma_i} &= \Gamma \left( \Sigma_i \rightarrow \bar {L}_{\alpha} \, H \right) + \Gamma \left( \Sigma_i \rightarrow L_{\alpha} \, \Phi \right) \nn \\ 
&=\frac{3}{32 \pi }\, M_{\Sigma_i} \, \left[ \left( Y_{H}^\dagger \, Y_H \right)_{ii} + \frac 43\, \left( Y_{\Phi}^\dagger \, Y_\Phi \right)_{ii} \right] \,, 
\end{align}
and $\Gamma_{\bar{\Sigma_i}}=\Gamma_{\Sigma_i}$. The branching ratios are given by
\begin{align}
    \text{Br}_\Phi^i &= \frac{4\, \left( Y_{\Phi}^\dagger \, Y_\Phi \right)_{ii}}{ 3\left( Y_{H}^\dagger \, Y_H \right)_{ii} + 4\, \left( Y_{\Phi}^\dagger \, Y_\Phi \right)_{ii}}\,,\n\\
    \text{Br}_H^i &= \frac{3\, \left( Y_{H}^\dagger \, Y_H \right)_{ii}}{ 3\left( Y_{H}^\dagger \, Y_H \right)_{ii} + 4\, \left( Y_{\Phi}^\dagger \, Y_\Phi \right)_{ii}}\,.
\end{align}
If the Yukawas are similar, the branching ratios become $\text{Br}_{\Phi} \simeq 0.6$ and $\text{Br}_H \simeq 0.4$.

The thermally averaged RIS-subtracted rate for $\Delta L=2$ scattering process in the $s$-channel is 
\begin{align}
    \gamma_{s}\left(\bar{L}_{\beta}H \to L_{\alpha} \Phi \right) =& \frac{3\, M_{\Sigma_1}^4}{1024\pi^5\,x}\,\left|Y_{\Phi_{\alpha 1 }} \right|^2 \, \left|Y_{H_{\beta 1 } }\right|^2\, \int dy \,y^{3/2}\n\\
    &\times \left[ \frac{\left(y -1 \right)^2 - a_{\Gamma}}{\left[ \left(y-1\right)^2 +a_\Gamma\right]^2}\right] \,K_1 \left(\sqrt{y}~x \right) \,, 
\end{align}
where we have defined $y = s/M_\Sigma^2 \,$ and $a_\Gamma = \Gamma^2/M_\Sigma^2$. On the other hand, for the $t$-channel we have
\begin{align}
   \gamma_t \left(\bar{L}_{\beta}H \to L_{\alpha} \Phi \right) =&\frac{ 3 \,M_{\Sigma_1}^4}{512\,\pi^5\,x}\, \left|Y_{\Phi_{\beta 1 }} \right|^2 \, \left|Y_{H_{\alpha 1 }} \right|^2\,\int dy \,  y^{-1/2}\n\\
   &\times \left( y -\ln\left( 1+y\right)\right)
K_1 \left( \sqrt{y}~ x  \right).
\end{align}

\section{Sphaleron Redistribution Coefficient}\label{app:Sphaleron}
The asymmetry created in the standard model leptons can be redistributed among the other fermions and scalars via the Yukawa interactions that are in equilibrium. These processes also known as spectator processes can lead to the conservation laws for the individual number densities, $n_{x}$ which for ultra relativistic bosons and fermions is given by 
\beqa \label{eq:ultra-rel}
n_{x} - n_{\bar x} & = & \frac{\mu_x \, g_x}{3}\, T^2 \begin{cases}
    \frac 12 & \text{for fermions} \\
    1 & \text{for bosons}
\end{cases},
\eeqa
where $\mu_x$ and $g_x$ stand for chemical potential and internal degrees of freedom of the species $x$, 
whereas for a massive particle the number density is expressed as 
\beqa \label{eq:non-rel}
n_{x} - n_{\bar x} & = & \frac{\mu_x \, g_x}{\pi^2}\, T^2 F_\pm \left( \frac{m_x}{T}\right)\,,
\eeqa
with
\begin{align}
   F_{\pm} = \int_{x}^{\infty} dy \frac{y\sqrt{y^2 -z^2} \, e^y}{\left( 1\pm e^y\right)^2} \, .
\end{align}
For our model the appropriate conservation laws read as: 
\begin{itemize}
    \item Hypercharge neutrality of the plasma:
        \beqa
        3\left( \mu_q +2\mu_u-\mu_d-\mu_\ell-\mu_e\right) + 2\mu_H +12\mu_\Phi = 0.
        \eeqa
    \item $SU(2)_L$ sphalerons: 
        \beqa
        3\mu_q +\mu_\ell =0 \quad \text{from  } {\cal O}_{\rm sph.} =  \Pi_{i=1}^{3} \ell_i q_i q_i q_i\, .
        \eeqa
    \item $SU(3)_c$ sphalerons: 
        \beqa
        2\mu_q-\mu_u-\mu_d = 0.
        \eeqa
    \item Yukawa interactions:
        \beqa
        \mu_\ell -\mu_H -\mu_e &=& 0 \quad \text{from} \quad \bar L He_R \,,\n \\
        \mu_q-\mu_H-\mu_d &=& 0 \quad \text{from} \quad \bar Q_L H d_R \, ,\n \\
        \mu_q+\mu_H-\mu_u & =& 0 \quad \text{from} \quad \bar Q_L \tilde{H} u_R \, .
        \eeqa
    \item Scalar interaction: 
        \beqa
        \mu_\Phi -3\mu_H = 0 \quad \text{from} \quad H^3\, \Phi^\ast\,.
        \eeqa
\end{itemize}
Here $q,\ell$ stand for the $SU(2)$ quark and lepton doublets, while $u,d,e$ stand for the right-handed up, down and electron singlets in the SM. Solving the above system of equations and defining 
\beqa
\mu_B = 3\times \left( 2\, \mu_q + \mu_u + \mu_d \right) \, , \nn \\
\mu_L = 3\times \left( 2\,\mu_\ell + \mu_e \right) -8\, \mu_\Phi ,\label{eq:app-BL}
\eeqa
we find that 
\beqa
\mu_B & = & \frac{100}{217} \, \mu_{B-L} \, , \nn \\
{\rm or},\quad \frac{\mu_B}{ \mu_{B-L}} & =& c_{\rm sph} \, \simeq \, 0.461\,. 
\eeqa
This is the sphaleron conversion factor used in our calculation of the final baryon asymmetry.

\bibliographystyle{JHEP}
\bibliography{refs}
\end{document}